\documentstyle[aps,prb,multicol,graphics]{revtex}

\begin{document}
\draft
\title{Dynamic critical exponent of two-, three-, and four-dimensional 
$XY$ models with relaxational and resistively shunted junction dynamics}
\author {Lars Melwyn Jensen, Beom Jun Kim, and Petter Minnhagen}
\address {Department of Theoretical Physics,
Ume{\aa} University, 901 87 Ume{\aa}, Sweden}
\preprint{\today}
\maketitle
\begin{abstract}
The dynamic critical exponent $z$ is determined numerically for 
the $d$-dimensional $XY$ model ($d=2, 3$, and 4)
subject to relaxational dynamics and resistively shunted junction dynamics.
We investigate both the equilibrium fluctuation and the relaxation behavior 
from nonequilibrium towards equilibrium, using the finite-size scaling method.   
The resulting values of $z$ are shown to depend on the boundary conditions 
used, the periodic boundary condition, and fluctuating twist boundary
condition (FTBC), which implies that the different treatments of the
boundary in some cases give rise to different critical dynamics. 
It is also found that the equilibrium scaling and the approach to equilibrium
scaling for the the same boundary condition do not always
give the same value of $z$.
The FTBC in conjunction with the finite-size
scaling of the linear resistance for both type of dynamics yields values
of $z$ consistent with expectations for superfluids and superconductors:
$z = 2$, 3/2, and 2 for $d=2$, 3, and 4, respectively.
\end{abstract}

\begin{multicols}{2}
\section{Introduction} \label{sec:intro}
Superconducting films, Josephson junction arrays, and superfluid $^4$He are systems
where topological defects play an important role close to the phase
transition. This is particularly striking in two dimensions (2D) where
a phase transition of the Kosterlitz-Thouless (KT) nature is driven by 
the unbinding of thermally created topological defects, vortex-antivortex 
pairs.~\cite{2Ddefect,minnhagen:rev}
In 3D such topological defects take the form of vortex loops and it has
been argued that the physics close to the transition
can be associated with these loops.~\cite{3Ddefect}
The common feature
in these systems is that they can be characterized by a complex order
parameter. The $XY$ model can be viewed as a discretized version of
such systems where only the phase of the complex order parameter plays a
significant role. This model is believed to catch the essential
features of  the topological defects present in $^4$He as well as
in superconductors in the limit when the magnetic penetration length
is much larger than the correlation length; high-$T_c$ superconductors
fall into this category.~\cite{minnhagen:fermi}
All the systems which can be described by the $XY$ model
belong to the same universality class for the thermodynamic
critical properties of the phase transition.

In the present paper we have the connection between the $XY$
model and superfluid and superconducting systems in mind. However, the $XY$
model {\it per se} can equally well be viewed as a simple model of a ferromagnet where
the phase angle corresponds to the direction of a 2D spin vector
associated with each lattice site.

Our interest in the present paper is the dynamical properties
associated with topological defects, which may of
course depend on the explicit choice of the dynamics imposed on the model.
We here investigate two types of dynamics: One is a simple
relaxational dynamics (RD) and the other is the resistively shunted 
junction dynamics (RSJD).
We calculate the dynamic critical exponent $z$ using various scaling
relations both associated with equilibrium and with the approach to
equilibrium when starting from a nonequilibrium configuration. 
Our main conclusion is that the dynamic critical exponents associated 
with the topological defects are the same for these two types of dynamics, RD and RSJD. 
However, this conclusion does depend on the precise treatment of the boundary.
We demonstrate that various values of $z$ can be obtained by changing
the treatment of the boundary, as well as by changing from scaling in
equilibrium to scaling for the approach to equilibrium.

This paper is organized as follows:
In Sec.~\ref{sec:model} we briefly introduce the $XY$ model and explain how
the dynamic equations are defined in RSJD and RD taking boundary conditions into account. 
Section~\ref{sec:scaling} describes the various scaling relations used to
obtain $z$. The results from our simulations are given in Sec.~\ref{sec:result}
for spatial dimensions $d=2$, 3, and 4, whereas Sec.~\ref{sec:conc} 
contains discussions of the results. Finally
Sec.~\ref{sec:summary} gives a short summary of the main conclusions.

\section{$XY$ Model and Dynamics} \label{sec:model}

\subsection{$XY$ model}
The $d$-dimensional $XY$ Hamiltonian on a hypercubic lattice of the
size $\Omega \equiv L^d$ is  defined by 
\begin{equation} \label{eq:H_XY}
H[\theta_{\bf r}] =-J\sum _{\langle \bf{rr}^{\prime }\rangle }\cos (\phi _{\bf{rr}^{\prime }} 
\equiv \theta_{\bf{r}} - \theta_{\bf{r}^\prime}),
\end{equation}
where the summation is over nearest neighboring pairs, $\theta_{\bf r}$
is the phase of the complex order parameter at position ${\bf r}$,
and $J$ is the coupling strength.  
The $XY$ Hamiltonian is appropriate not only to describe the overdamped
Josephson junctions arrays without charging energy, but  can also
be viewed as a discretized form of the Ginzburg-Landau (GL) free energy 
\begin{equation} \label{eq:GL}
F_{GL}[\psi({\bf r})]=\int d{\bf r}\left( \alpha |\psi ({\bf r}) |^{2}+\frac{\beta }{2}|\psi ({\bf r})|^{2}+\frac{1}{2}|\nabla \psi ({\bf r})|^{2}\right) ,
\end{equation}
where the amplitude fluctuations of the complex order parameter $\psi({\bf r})$ are neglected: 
$\psi ({\bf r})=\psi_0e^{i\theta (\bf{r})}$ with $\psi _{0}$
fixed to a constant. When mapping the GL free energy functional onto the $XY$ Hamiltonian
the coupling strength $J$ is found to be proportional to $|\psi_0|^2$.

The thermodynamic properties of the $XY$ model have been intensely studied for
many years and it is well known that the important length scale in the critical region, 
the correlation length $\xi$, diverges at the critical temperature $T_c$.  
In 3D and 4D the divergence is of the standard form of the continuous second-order
transition, i.e., $\xi(T)\sim|T-T_c|^{-\nu}$,  whereas in 2D $\ln \xi(T) \sim(T-T_c)^{-1/2}$ 
as $T_c$ is approached from above and $\xi=\infty$ in the whole
low-temperature phase where quasi-long-range order exits in the absence of true long-range
order.~\cite{2Ddefect} From the point of view of the finite-size scaling, this 
feature of the 2D KT transition turns the finite system size $L$ into the relevant 
length scale in the low-temperature phase.

\subsection{Boundary condition} \label{subsec:bc}
Experiments on superconductors and $^4$He are usually done on samples with open
boundaries. From this perspective it is preferable to use
boundary conditions, which reflects this experimental situation also in
the simulations. However, simulations of the $XY$ model can usually
only be well converged on relatively small lattice sizes, and 
since the surface to volume ratio is inversely proportional to the
linear system size $L$, the open boundary gives rise to large surface
effects, which decay very slowly as the system size is increased. 
The standard way of reducing these unwanted surface effects 
is to impose the periodic boundary condition (PBC): $\theta_{ {\bf r}+L\hat{ \bf \mu }}=
\theta_{ \bf r}$, where $\hat{ \bf \mu}$
denotes the basis vectors of the lattice, 
e.g., $\hat{\mu} = \hat x, \hat y, \hat z$ in 3D. One drawback of this boundary
condition is that it restricts the twist from ${\bf r}$ to ${\bf r}+L\hat{\mu}$, 
defined as the sum of the phase differences along a direct path connecting
the two positions, to an integer multiple of $2\pi$. On the
other hand, this twist from one boundary to the opposite for an open system 
can have any value. It is thus preferable to relax the PBC so as to 
allow for a continuous twist by changing the boundary condition to a more
generalized form: $\theta_{ {\bf r}+L\hat{\mu }}=\theta_{\bf r}+L\Delta_{\mu }$,
which has been used in various contexts.~\cite{eikman90,alvarez97,olsson,bjkim99a}
In particular the boundary condition where the twist variable $\Delta_{\mu }$ is not fixed
to a constant but allowed to 
fluctuate has been termed the fluctuating twist boundary condition 
(FTBC), which was originally introduced for static 
Monte Carlo (MC) simulations~\cite{olsson} and then extended 
to Langevin-type dynamics at finite temperatures.~\cite{bjkim99a}
Since the FTBC allows for any value of the twist, it is closer to the 
open boundary condition for a real system.
Of course one does not expect the treatment of the
boundary to affect the results in the thermodynamic limit. However,
as we will show and discuss here, the dynamics at criticality can
depend on the boundary condition, in as far as the dynamic critical
exponent can be defined in terms of the finite-size scaling.
It is worth mentioning that a similar 
observation, i.e.,  that an important exponent may depend on boundary conditions,
has been made recently in the study of the stiffness exponent of
vortex-glass models.~\cite{kosterlitz99}

\subsection{Dynamic models} \label{subsec:dynamics}
Next we introduce two simple dynamic models 
widely used to describe behaviors of superfluids, superconducting
films, regular Josephson junction arrays, and also bulk high $T_c$ superconductors
close to the transition temperature. 

\subsubsection{Resistively shunted junction dynamics} \label{subsubsec:RSJD}
A $d$-dimensional hypercubic array of size $\Omega =L^d$ ($L=$ linear size)
of superconducting grains weakly coupled by resistively shunted
Josephson junctions is effectively described by the $XY$ Hamiltonian (\ref{eq:H_XY})
when it comes to the static properties. On the other hand, dynamic equations of motion for 
the corresponding overdamped
RSJ model are generated from local conservation of the current on each grain. 
The total current $I_{\bf{rr}^\prime}$ between neighboring grains (${\bf r}, {\bf r}^\prime$)
is the sum of the supercurrent, the normal resistive current, and the thermal noise current: 
$I_{\bf{rr}^\prime}  = I^{s}_{\bf{rr}^{\prime }}+I^{n}_{\bf{rr}^{\prime }}+
I^t_{\bf{rr}^{\prime }}$. 
The supercurrent is given by the Josephson current-phase relation
$I^{s}_{\bf{rr}^{\prime}}=I_c\sin (\phi _{\bf{rr}^{\prime }})$,
where $I_c=2eJ/\hbar$ is the critical current for a single junction.
The normal resistive current $I^{n}_{\bf{rr}^{\prime}}=V_{\bf{rr}^{\prime }}/R_0$,
where the voltage difference $V_{\bf{rr}^{\prime }}$ is related to the phase difference 
by $V_{\bf{rr}^{\prime}} = (\hbar/2e)\dot\phi_{\bf{rr}^{\prime}}$ and  $R_0$ is
the shunt resistance. Finally the thermal noise currents $I^t_{\bf{rr}^\prime}$ in the shunts
satisfy $\langle I^t_{\bf{rr}^\prime } \rangle = 0$
and 
\begin{equation}
\langle I^t_{ {\bf r}_1 {\bf r}_2 }(t)I^t_{ {\bf r}_3 {\bf r}_4}(0)\rangle
=\frac{2k_BT}{R_0}\delta(t)\left( 
\delta_{ {\bf r}_1 {\bf r}_3} \delta_{ {\bf r}_2 {\bf r}_4}
-\delta_{ {\bf r}_1 {\bf r}_4} \delta_{ {\bf r}_2 {\bf r}_3}\right), 
\end{equation}
where $\langle \cdots \rangle$ is the thermal average, and $\delta(t)$ and 
$\delta_{ {\bf r} {\bf r}'}$ are the Dirac and Kronecker deltas, respectively.
From local current conservation we obtain
\begin{equation}
\sum _{\hat{n}}I_{ {\bf rr}+\hat{n}}=I^{\rm ext}_{\bf{r}},
\end{equation}
where the $\hat{n}$ summation is over $2d$ nearest neighbors of site ${\bf r}$
on a hypercubic lattice in $d$ dimensions ($\hat{n}=\pm \hat{\mu}$), e.g., 
$\hat{n} = \pm \hat x, \pm \hat y, \pm \hat z$ in 3D,
and $I_{\bf{r}}^{\rm ext}$ is an external current source at $\bf{r}$
(in the present work, we only consider the case without external driving:
$I^{\rm ext}_{\bf r} = 0$).
Introducing the lattice Green's function $U_{\bf{rr}^{\prime}}$, which
is the inverse of the discrete Laplacian, 
the RSJD equations of motion in the absence of external currents can be written in dimensionless form as
\begin{equation} \label{eqn:theta}
\frac{d\theta _{\bf{r}}}{dt}=-\sum _{\bf{r}^{\prime }}\bar{U}_{\bf{rr}^{\prime }}\sum _{\hat{n}}
\sin (\theta_{{\bf r}^{\prime }} - \theta_{{\bf r}^{\prime }+\hat{n}} )+\zeta_{\bf{r}},
\end{equation}
where $\bar{U}_{\bf{rr}^{\prime }} \equiv U_{\bf{rr}^{\prime }}-U_{\bf{rr}}$,~\cite{shenoy85b}
and from here on we normalize the time, the current, the distance, the energy, and the 
temperature in units of $\hbar/2eR_0 I_c$, $I_c$, the lattice spacing $a$, $J$,
and $J/k_B$, respectively.
The on-site noise term $\zeta_{\bf r}(t) \equiv -\sum _{{\bf r}^\prime}
\bar{U}_{\bf{rr}^{\prime }}\sum_{\hat{n}}I^t_{{\bf r}^{\prime }{\bf r}^{\prime }+
\hat{n}}(t)$ is spatially correlated, which is a consequence of
the local current conservation, and satisfies $\langle \zeta_{\bf{r}}(t)
\zeta_{\bf{r}^{\prime }}(0)\rangle = 2T\bar{U}_{\bf{rr}^{\prime }}\delta(t)$.
The RSJD equations~(\ref{eqn:theta}) can be rewritten 
in a Langevin-type form~\cite{shenoy85b}
\begin{equation}
\label{eqn:rsj_langevin}
\frac{d\theta _{\bf{r}}}{dt}=-\sum _{\bf{r}^{\prime }}\bar{U}_{\bf{rr}^{\prime }}\frac{\delta 
H[\theta _{\bf{r}}]}{\delta \theta _{\bf{r}^{\prime }}(t)}+\zeta _{\bf{r}} ,
\end{equation}
with the $XY$ Hamiltonian $H$ in Eq.~(\ref{eq:H_XY})
[compare with Eq.~(\ref{eqn:tdgl_theta}) for relaxational dynamics].

We now introduce the FTBC for the RSJD.
The global twist $L\Delta_\mu$ in the $\hat \mu$ direction across the whole system
(see Sec.~\ref{subsec:bc}) is introduced through the local transformation
$\theta _{\bf{r}} \to \theta _{\bf r}+{\bf r}\cdot {\bf \Delta }$, 
still keeping $\theta _{\bf r}=\theta_{ {\bf r}+L\hat{\mu }}$ as
the periodic part of the phases. The Hamiltonian in terms of these variables 
is 
\begin{equation} \label{eq:twisted_H}
H[\theta_{\bf r}, {\bf \Delta}]=-\sum _{\bf{r}\hat{\mu }}\cos(\theta_{\bf r}-
\theta_{{\bf r}+\hat{\mu}}-\hat{\mu}\cdot {\bf \Delta}),
\end{equation}
where the $\hat{\mu}$ summation is over $d$ nearest neighbors of ${\bf r}$
in each positive direction (e.g., ${\hat \mu} = \hat x, \hat y, \hat z$ in 3D).
It is straightforward to show that equations of motion for phase variable $\theta_{\bf r}$
are given by Eq.~(\ref{eqn:rsj_langevin}) with the substitution $H$
given by Eq.~(\ref{eq:twisted_H}):~\cite{bjkim99a}
\begin{equation}
\label{eqn:rsj_langevin_twisted}
\frac{d\theta _{\bf{r}}}{dt}=-\sum _{\bf{r}^{\prime }}\bar{U}_{\bf{rr}^{\prime }}\frac{\delta
H[\theta_{\bf r}, {\bf \Delta}]}{\delta \theta _{\bf{r}^{\prime }}(t)}+\zeta _{\bf{r}} .
\end{equation}
In order to get a closed set of equations we further have to specify the dynamics of the twist
variables $\Delta_{\mu}$, which is simply the average phase difference between 
opposite faces on the $d$-dimensional hypercube.
In the absence of external currents, the physical boundary condition,
corresponding to an open boundary in real systems, 
should satisfy the condition that there be no current across the boundary,
which leads to~\cite{bjkim99a}  
\begin{equation}
\frac{d\Delta _{\mu }}{dt}=\Gamma_{\Delta }\sum_{\bf r}
\sin (\theta _{\bf{r}}-\theta _{\bf{r}+\hat{\mu }}-\Delta _{\mu })+\zeta ^{\Delta }_{\mu }
\end{equation}
or, equivalently,
\begin{equation} \label{eqn:delta}
\frac{d\Delta_\mu}{dt}= - \Gamma_\Delta
\frac{\delta H[\theta_{\bf r}, {\bf \Delta} ]}{\delta\Delta_\mu}+\zeta_\mu^\Delta, 
\end{equation}
where $\Gamma_\Delta=1/L^d$.
As shown in Ref.~\onlinecite{bjkim99a} the noise term satisfies 
$\langle\zeta^\Delta_\mu(t)\rangle = \langle\zeta^\Delta_\mu(t)\zeta_{\bf r}(t')\rangle = 0$ and 
$\langle\zeta_\mu^\Delta(t)\zeta_\nu^\Delta(0)\rangle
= 2T\Gamma_\Delta\delta_{\mu\nu}\delta(t)$. We term the dynamics defined in
this way [by Eqs.~(\ref{eqn:rsj_langevin_twisted}) and (\ref{eqn:delta})]
RSJD with the FTBC, whereas the RSJD with the PBC is given by 
Eq.~(\ref{eqn:rsj_langevin}) with $H$ in Eq.~(\ref{eq:H_XY}).

\subsubsection{Relaxational dynamics}
Next we introduce the simpler phenomenological relaxational dynamics
called time-dependent Ginzburg-Landau-Langevin dynamics, which
represents a nonconserved dynamics,~\cite{hohenberg77} 
for the complex order parameter $\psi_{\bf r}$
on a discrete lattice:
\begin{equation}
\frac{d\psi_{\bf r}}{dt} = -\Gamma \frac{\delta F_{GL}[\psi_{\bf r}]} 
{\delta \psi_{\bf r}(t)}+\zeta_{\bf r},
\end{equation}
where $\Gamma$ is the diffusion constant, $F_{GL}$ is the 
discrete version of the GL free energy functional~(\ref{eq:GL}),
and the white noise term satisfies
$\langle \zeta_{\bf{r}}(t)\rangle =0$
and $\langle \zeta_{\bf{r}}(t)\zeta _{\bf{r}^{\prime }}(0)\rangle =
2k_BT\Gamma \delta_{\bf{rr}^{\prime }}\delta (t)$.
The order parameter relaxes towards a
configuration which locally minimizes the free energy, and the noises force the metastable
states to decay. In the London limit the system can be described solely by the
phase \( \theta _{\bf{r}}(t) \) of the order parameter 
$\psi_{\bf r} = \psi _{0}e^{i\theta _{\bf{r}}}$ with $\psi_0$ fixed to a constant.
Hence, by neglecting the amplitude fluctuations and discretizing the time-dependent
Ginzburg-Landau equation of motion, we find the phase equations of motion for
the RD defined by 
\begin{equation}
\label{eqn:tdgl_theta}
\frac{d\theta _{\bf{r}}}{dt}=-\frac{\delta H[\theta _{\bf{r}}]}{\delta \theta_{\bf r}(t)}
+\zeta_{\bf r},
\end{equation}
where $H$ is the $XY$ Hamiltonian~(\ref{eq:H_XY}) in units of $J$, 
the time unit is $\hbar/\Gamma J$,  and the dimensionless thermal
noises satisfy $\langle \zeta_{\bf r}(t)\rangle   =  0$
and 
\begin{equation}
\langle \zeta_{\bf{r}}(t)\zeta_{\bf{r}^{\prime }}(0)\rangle   
=  2T\delta (t)\delta _{\bf{rr}^{\prime }} , 
\end{equation}
with $T$ in units of $J/k_B$.
From Eq.~(\ref{eqn:tdgl_theta}), the RD equations for the phases in the case of 
the PBC are given by
\begin{equation}
\label{eqn:rd_eom}
\frac{d\theta_{\bf{r}}}{dt}=-\sum _{\bf{r}\hat{n}}
\sin (\theta_{\bf r} - \theta_{ {\bf r}+\hat{n}})
+\zeta_{\bf r} , 
\end{equation}
with periodicity on the phase variables: $\theta_{\bf r} = \theta_{ {\bf r} + L{\hat \mu}}$.

We now proceed to the case of the FTBC for RD. In this case,  in addition
to the equations of motion for phases {\bf (}Eq.~(\ref{eqn:tdgl_theta}) with substitution
$H[\theta_{\bf r}]$ by $H[\theta_{\bf r}, {\bf \Delta}]$ in Eq.~(\ref{eq:twisted_H}) {\bf )}
we need dynamic equations for the twist variables $\Delta_\mu$. Relaxational dynamics
means that these equations are of the form
\begin{equation} 
\frac{d\Delta_\mu}{dt}= - \Gamma_\Delta
\frac{\delta H[\theta_{\bf r}, {\bf \Delta} ]}{\delta\Delta_\mu}+\zeta_\mu^\Delta,
\end{equation}
which is identical to the form derived for
RSJD [see Eq.~(\ref{eqn:delta}), where $\Gamma_\Delta = 1/L^d$ 
was determined from the requirement that no current flows through
the boundary]. We here define the dynamic equations for $\Delta_\mu$
in the RD case with the same value of $\Gamma_\Delta$, which makes the
equations identical to the corresponding equations in RSJD. 
Within the same interpretation that
$I^{n}_{ {\bf r}{\bf r} + {\hat\mu}}={\dot\phi}_{{\bf r}{\bf r}+\hat\mu}$
and $I^{s}_{ {\bf r}{\bf r} + {\hat\mu}}=\sin(\phi_{{\bf r}{\bf r}+\hat\mu})$
with $\phi_{{\bf r}{\bf r}+\hat\mu} = \theta_{\bf r} - \theta_{ {\bf r}^\prime } - \Delta_\mu$
as for RSJD (see Sec.~\ref{subsubsec:RSJD}), we are again imposing
a condition consistent with that there be no current across the boundary.

In the simulations, the coupled equations of motion are 
discretized in time (we use the discrete time step $\Delta t = 0.05$
and 0.01 for RSJD and RD, respectively) and numerically integrated using
the second order Runge-Kutta-Helfand-Greenside (RKHG) algorithm,~\cite{batrouni85}
which is much more efficient than the first-order Euler algorithm
since it can reduce the effective temperature shift~\cite{bjkim99a} due to the discrete 
time step significantly.
In the case of RSJD we apply the efficient fast Fourier transformation 
method (see, for example, Ref.~\onlinecite{eikman90}), which makes
the overall computing time $O(L^d \log_2 L)$ in $d$ dimensions.
[For comparison, the RD simulation requires $O(L^d)$.]
The thermal noises are generated from a uniform distribution, whose width
is determined to satisfy the noise correlation (see above) at a given
temperature.

\section{Scaling Relations} \label{sec:scaling}

\subsection{Scaling in equilibrium}
In order to obtain the dynamic critical exponent $z$ from equilibrium
fluctuations of the system we use two different scaling relations: 
One is the finite-size
scaling of the time correlations of the supercurrent and the other is
the finite-size scaling of the linear resistance.
Fisher {\it et al.},~\cite{fisher91} proposed a general scaling theory 
of the conductivity for a homogeneous superconductor, which has been
studied further explicitly by Dorsey and co-worker.~\cite{dorsey91,wickham99}
The predictions from this scaling theory are very general and  depend only
on the dynamic scaling
assumption and the existence of a diverging correlation length 
$\xi \sim |T-T_c|^{-\nu }$:
From a simple dimensional analysis, it is easy to show that the order 
parameter scales as $\psi \sim \xi^{1-d/2}$, and thus the superfluid density 
scales as $\rho_{s} \sim |\psi|^2 \sim \xi^{2-d}$. Below $T_c$
one has $\sigma (\omega )\sim i\rho _{s}/\omega$, and accordingly
one deduces that the frequency-dependent linear conductivity scales 
as~\cite{fisher91} 
\begin{equation}\label{FFH}
\sigma (\omega) = \xi^{2-d+z}F_{\sigma }\left (\omega \xi^{z}\right), 
\end{equation}  
where $F_{\sigma}$ is a universal scaling function, the
dynamic critical exponent $z$ is introduced from  $\tau \sim \xi^z$,
and $\tau$ is the characteristic time scale.
Precisely at $T_c$, Eq.~(\ref{FFH}) turns into the finite-size scaling 
form of the conductivity:
\begin{equation}
\sigma (\omega) =  L ^{2-d+z}F_{\sigma }\left (\omega L ^{z}\right).
\end{equation}
This scaling relation can be put to practical use in the case of the PBC
because for this boundary condition $\rho_s$ has the required size
scaling. On the other hand, it cannot be used for an open boundary condition or
for the FTBC because in these cases $\rho_s=0$ at any $L$ and $T$.~\cite{olsson} 
For the FTBC we will then instead use the finite-size scaling of the linear
resistance described below.

\subsubsection{Scaling of supercurrent correlations\label{subsubsec:supercurrent}}

The conductivity $\sigma (\omega)$ may be related to the 
supercurrent correlation function $G(t)$, which for the $XY$ model in $d$ dimensions
is given by
\begin{equation} \label{eq:G}
G(t) = \frac{1}{L^d}\left\langle F(t) F(0) \right\rangle, 
\end{equation} 
where the global supercurrent $F(t)$ flowing in a given direction, say, $\hat x$,
is written as 
\begin{equation} \label{eq:F}
F(t) = \sum_{\bf r} \sin(\theta_{\bf r} - \theta_{ {\bf r} + {\hat x} }).
\end{equation} 
The correlation function $G(t)$ is a key quantity in describing
the dynamic response of vortex fluctuations and is for $t=0$ directly related
to the static helicity modulus.~\cite{olsson} The connection between
$\sigma(\omega)$ and $G(t)$ in the RSJD case is expressed as~\cite{bjkim99a}
\begin{equation} \label{eq:sigma}
\sigma(\omega) = 1 + \frac{i \rho_s}{\omega} - \frac{1}{T}
\int_0^\infty dt e^{i\omega t} G(t), 
\end{equation}
where the conductivity is measured in units such that the shunt resistance $R_0= 1$, 
and the superfluid density $\rho_s$ is given by
\begin{equation}
\rho_s = \rho_0 \left( 1 - \frac{1}{\rho_0T} G(t = 0) \right) , 
\end{equation}
with the bare superfluid density 
$\rho_0 \equiv \langle \cos(\theta_{\bf r} - \theta_{ {\bf r} + {\hat x} }) \rangle$.
The dynamic dielectric function $1/\epsilon(\omega)$ in 2D  
is also expressed as~\cite{houlrik}
\begin{equation}
{\rm Re}\left[
\frac{1}{\epsilon(\omega)}\right]=\frac{1}{\epsilon(0)}+\frac{\omega}{\rho_0T}
\int_0^\infty dt \sin \omega t G(t), 
\end{equation}
\begin{equation}
{\rm Im}\left[
\frac{1}{\epsilon(\omega)}\right]= - \frac{\omega}{\rho_0T}
\int_0^\infty dt \cos \omega t G(t), 
\end{equation}
where
\begin{equation} \label{eq:epsilon0}
  \frac{1}{\epsilon (0)}=1-\frac{1}{\rho_0T}G(0).
\end{equation}
The helicity modulus $\gamma$ corresponds to the superfluid density $\rho_s$ and
is given by $\gamma = \rho_s=\rho_0/\epsilon(0)$. The conductivity $\sigma(\omega)$
in RSJD can be further simplified into the form~\cite{bjkim99a}
\begin{equation}\label{sigma-eps}
  \sigma (\omega)=1-\frac{1}{i\omega}\frac{\rho_0}{\epsilon(\omega )} .
\end{equation}
Expressing the scaling in terms of $G(t)$ leads to the scaling form

\begin{equation} \label{xiscale}
  G(t)=\xi^{2-d}F_G(t\xi^{-z}), 
\end{equation}
which at $T_c$ for 3D turns into the finite-size scaling form
(see Appendix~\ref{append:scale})
\begin{equation} \label{Lscale_3D}
  LG(t)=F_G(tL^{-z}), 
\end{equation}
while in 2D a logarithmic correction (see Appendix~\ref{append:scale})
needs to be included 
\begin{equation}\label{Lscale_2D}
\ln\left(\frac{L}{c}\right) G(t)= F_G(tL^{-z}),
\end{equation}
where $F_G(x)$ is the scaling function for $G(t)$.
In the following we will use the scaling relations Eqs.~(\ref{xiscale})
and (\ref{Lscale_3D}) in 3D with the PBC and
Eq.~(\ref{Lscale_2D}) in 2D with the PBC.

\subsubsection{Resistance scaling\label{subsubsec:resistance} }
In order to obtain a finite-size scaling at criticality for the FTBC for which,
like for any open boundary condition, $\rho_s=0$ at any temperature
and any lattice size, we relate the resistance $R$ to the fluctuations of the
twist over the sample. The voltage across
the sample in the $\hat\mu$ direction $V_\mu = -L\dot{\Delta}_\mu$ (see Ref.~\onlinecite{bjkim99a})
and the linear resistance $R_\mu$ in the same
direction are related to the voltage fluctuation by the fluctuation-dissipation 
theorem~\cite{reif} 
\begin{eqnarray} 
R_\mu & = & \frac{1}{2T} \int_{-\infty}^\infty dt \langle V_\mu(t) V_\mu(0) \rangle \label{eq:R1} \\
&\approx & \frac{L^2}{2 T } \frac{1}{\Theta} \left \langle [ 
\Delta_\mu(\Theta) - \Delta_\mu(0) ]^2
\right\rangle , \label{eq:Rmu}
\end{eqnarray}
where the approximation becomes exact for a sufficiently large 
time $\Theta$, as shown in Appendix~\ref{append:R} (a similar approximation has
been used for RSJD with open boundary condition in Ref.~\onlinecite{granato}).
In the present simulation we use $\Theta=2000$ and perform average over 
all $d$ directions, i.e., $R=(\sum_\mu R_\mu)/d$. 

Since $R_\mu$ scales as the inverse of the characteristic time scale
in the critical region, the finite-size scaling takes the form~\cite{lidmar98}
\begin{equation}\label{rscale_full}
R=\frac{1}{L^z}F_R\bigl( (T-T_c)L^{1/\nu} \bigr), 
\end{equation}
where $\nu$ is the critical exponent for correlation length ($\xi \sim |T-T_c|^{-\nu}$)
and $F_R(x)$ is the scaling function for $R$.
Precisely at $T_c$, $F_R(x) = F_R(0)$ becomes a constant independent of $L$
and we get  
\begin{equation}\label{rscale}
  R\sim L^{-z}, 
\end{equation}
which can be used to determine $z$, once $T_c$ is known. 
The resistance scaling can also be turned into an {\it intersection method}~\cite{lidmar98}
for determining $z$ and $T_c$ using that
\end{multicols}
\noindent\rule{0.5\textwidth}{0.1ex}\rule{0.1ex}{2ex}\hfill
\begin{equation}\label{inter}
  \frac{\ln(R_L/R_{L^\prime})}{\ln(L/L^\prime)} = -z
 + \frac{ \ln \left[ F_R\bigl( (T-T_c)L^{1/\nu}\bigr) /
     F_R\bigl( (T-T_c){L^\prime}^{1/\nu} \bigr) \right]} {\ln(L/L^\prime)},
\end{equation} 
\hfill\raisebox{-1.9ex}{\rule{0.1ex}{2ex}}\rule{0.5\textwidth}{0.1ex}
\begin{multicols}{2}
\noindent
for two different lattice sizes $L,L^\prime$. 
Thus, if we plot $\ln(R_L/R_{L^\prime})/\ln(L/L^\prime)$ as 
a function of temperature for several pairs of sizes ($L, L^\prime$), 
all curves intersect at a single unique point $(T_c,-z)$.~\cite{lidmar98}
Once $T_c$ and $z$ are determined through the above intersection method, all data
can be made to collapse onto a single scaling curve by plotting $RL^z$ as a function 
of the scaling variable $(T-T_c)L^{1/\nu}$ with the correct value of the 
exponent $\nu$ [see Eq.~(\ref{rscale_full})].

\subsection{Scaling of relaxation towards equilibrium: Short-time relaxation}
\label{subsec:shorttime}
Recently, it has been found that a universal scaling in time 
can also be constructed for the relaxation towards equilibrium when starting
from a nonequilibrium configuration. Since such a relaxation is usually rather
fast, it is often referred as the short-time relaxation method.~\cite{li}
By this method several
critical exponents have been successfully determined for the unfrustrated
and the fully frustrated Josephson junction array~\cite{luo97} as well as for the Ising
model.~\cite{li,soares} In these studies Glauber dynamics in MC
simulations has been used to obtain time series of measured quantities, such as the
magnetization and the Binder's cumulant.
Here we apply this method to the $XY$ models 
with more realistic dynamics, RSJD and RD, both introduced in 
Sec.~\ref{subsec:dynamics}, in order to determine the value of the dynamic 
critical exponent $z$.
For convenience we measure~\cite{soares}
\begin{equation}
\tilde\psi  = \left \langle {\rm sign}
\left[ \sum_{\bf r} \cos\theta_{\bf r}(t)\right] \right \rangle ,
\end{equation}
starting from the initial condition $\theta_{\bf r}(0) = 0$.
Since $\tilde\psi(t=0) = 1$ at any system size $L$, the finite-size
scaling form becomes
\begin{equation} \label{eq:short}
\tilde\psi(t,T,L) = F_\psi\bigl(t/L^z,(T-T_c)L^{1/\nu}\bigr) ,
\end{equation}
with the scaling function $F_\psi(x,y)$ depending on two scaling variables, 
satisfying $F_\psi(0,y) = 1$ at any $y$. At $T_c$, $z$ is easily determined from 
Eq.~(\ref{eq:short}) because in this case the second argument
of the scaling function vanishes and $\tilde\psi(t)$ curves obtained for different sizes 
can be collapsed onto a single curve when plotted against the variable $t/L^z$.
We can also determine $T_c$ by an intersection method similar to Eq.~(\ref{inter}) 
as follows:
If the first argument of the scaling function is fixed to a constant ($t/L^z = a$)
for a given system size $L$ and $z$, then $\tilde\psi$ has
only one scaling variable $(T-T_c)L^{1/\nu}$, and can thus be written as 
\begin{equation} \label{eq:short_a}
\tilde{\psi}= {F}_\psi\bigl(a, (T-T_c)L^{1/\nu}\bigr).  
\end{equation}
Accordingly, if we plot $\tilde\psi$ with fixed $a$ as a function 
of $T$ for various $L$, all curves should intersect at $T_c$. 
However, because $a$ depends on the value of $z$ which cannot be independently 
determined by this method we start the intersection method from
the $z$ value determined from the scaling collapse at $T_c$.
The values of $T_c$ and $z$ obtained from this intersection method can be refined by 
the iterating intersection construction. Finally, to examine the consistency we 
collapse the data for all temperatures and lattice sizes onto a single scaling 
curve in the variable $(T-T_c)L^{1/\nu}$ 
at fixed $a = t/L^z$, which in addition is a check of the consistency 
against  the known value of the static exponent $\nu$.

\section{Simulation Results} \label{sec:result}
\subsection{2D $XY$ model} \label{subsec:result:2d}

In two dimensions, there has been some controversy over the value
of the dynamic critical exponent: There has been a theoretical
approach by Ambegaokar, Halperin, Nelson,
and Siggia~\cite{ahns} (AHNS) predicting 
$z_{\rm AHNS} = 1/2\tilde\epsilon T^{\rm CG}$, where the Coulomb gas (CG)
temperature $T^{\rm CG} \equiv T/2\pi\rho_0$ and $1/\tilde{\epsilon}\equiv 1/\epsilon(0)$
(see Sec.~\ref{subsubsec:supercurrent}).
On the other hand, a simple scaling 
argument has yielded $z_{\rm scale} = 
1/\tilde{\epsilon}T^{\rm CG}-2$.~\cite{PRL-z}
Also, in numerical simulations, there have been some differences: On the one hand, $z_{\rm AHNS}$
has been observed in Ref.~\onlinecite{simkin} from RSJ simulations, 
while $z_{\rm scale}$ has been concluded for RSJD and RD
(Refs.~\onlinecite{bjkim99a} and ~\onlinecite{PRL-z}), 
for Langevin dynamics of CG gas particles (Ref.~\onlinecite{kenneth}),
and for the MC simulation of lattice CG (Ref.~\onlinecite{weber97}).
Although the question is not completely resolved yet, 
we strongly believe that when the fluctuating twist boundary 
condition~\cite{bjkim99a}
(see Ref.~\onlinecite{gun} for a comparison between a conventional
boundary condition and the FTBC) is used, $z_{\rm scale}$
is the correct result.
Although the above mentioned two $z$ values are different below
the KT transition, they give the same value of 2 at the KT transition. 
In Ref.~\onlinecite{tiesinga}, however, $z \approx 1$ was concluded
from a simulation of RSJ dynamics with the PBC, while in Refs.~\onlinecite{pierson} 
and \onlinecite{dierk} a  very large value $z \approx 5$ has been suggested 
from a scaling analysis of existing 
experimental data and from an analytic calculation using Mori's technique, respectively.

In the low-temperature phase of the 2D $XY$ model,
we can alternatively derive $z_{\rm scale}$ in the following way:
The potential barrier, which a bound vortex-antivortex pair
should overcome in order to escape, is given by
\[
\Delta V = \frac{T}{{\tilde \epsilon}T^{\rm CG}} \ln L , 
\]
and the escape ratio $\Gamma \sim \exp(-\Delta V / T)$ for one pair is simply
related to the total probability of escape, $P$, by
\[ P = L^2 n \Gamma, \]
where $n$ is the vortex pair density.
The time scale $\tau$ of the system is inversely proportional to $P$
and thus is given by
\[
\tau \sim \frac{\exp(\Delta V/T)}{L^2 n} \sim L^{1/{\tilde \epsilon} 
  T^{\rm CG} - 2} \sim L^z
\]
and we obtain the dynamic critical exponent 
\[ z= \frac{1}{{\tilde \epsilon} T^{\rm CG}}  - 2\]
in accordance with Ref.~\onlinecite{PRL-z}, where $z$ has been obtained from
a simple scaling argument and the observed $1/t$ behavior of the correlation
function $G(t)$.

In this section, we investigate the dynamic critical exponent of the 2D $XY$ model
with RSJD and RD at, below, and above the KT transition.
We use the FTBC as well as the conventional PBC and use various methods such as the resistance scaling, the scaling of
the supercurrent correlation function, and the short-time relaxation method.
The results are summarized in Table~\ref{table:2d}.
As seen from Table~\ref{table:2d} only the FTBC gives results in accordance with the
expected value: $z_{\rm scale}=z_{\rm AHNS}\approx 2$ 
at $T=0.90 (\approx T_c)$,~\cite{olsson:Tc} whereas $z_{\rm scale}\approx 3.4$ and $z_{\rm AHNS}
\approx 2.8$ at $T=0.80$.~\cite{bjkim99a}
Furthermore, this is the case both for RD and RSJD. In contrast, the results for the PBC are
inconsistent both with $z_{\rm scale}$ and $z_{\rm AHNS}$. From this we
conclude that the FTBC is an adequate boundary condition in the context of
open systems like superfluid and superconducting films. It is also
interesting to note that also the short-time relaxation method for
RSJD with the FTBC gives results consistent with $z_{\rm
  scale}$. The results in Table~\ref{table:2d} will be further discussed in Sec.~\ref{sec:conc}.
In the following we present the simulation results on which 
Table~\ref{table:2d} is based. 

\subsubsection{Critical temperature} 
\begin{figure}
\centering{\resizebox*{!}{7.5cm}{\includegraphics{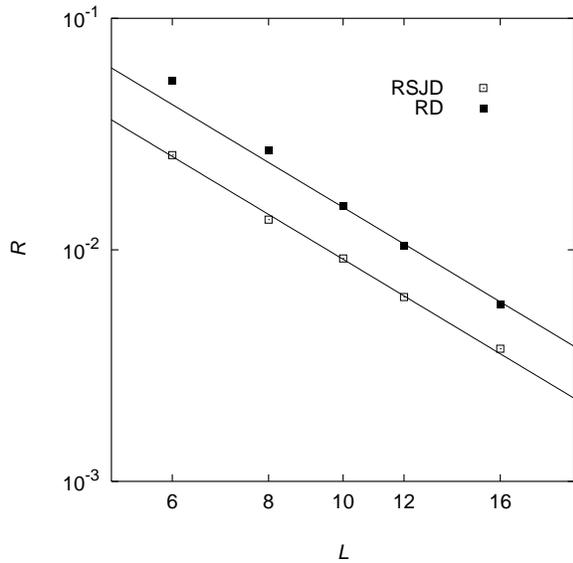}}}
\caption{Resistance scaling for 2D RSJD (open squares) and RD (solid squares) 
with the FTBC at $T$=0.9. The solid lines represent $R \sim L^{-2.0}$.
For both dynamics, $z \approx 2.0$ is obtained. (The data points were
taken from Ref.~\protect\onlinecite{bjkim99a}.)}
\label{fig:2d_R_Tc}
\end{figure}

First we fix the temperature to $T=0.90 \approx T_c$ and
focus on the dynamic critical behaviors at the KT transition.
The results from the resistance scaling $R\propto L^{-z}$ for the FTBC (see 
Sec.~\ref{subsubsec:resistance}) are displayed in Fig.~\ref{fig:2d_R_Tc} 
(the data points are taken from Ref.~\onlinecite{bjkim99a}), 
where the slopes of the lines in the log-log plot correspond
to $z\approx 2$  for both RSJD and RD. Consequently, our result $z \approx 2.0$ is in accordance with other existing theoretical predictions~\cite{ahns,PRL-z} while it 
contradicts recently suggested, very large values in Refs.~\onlinecite{pierson}
and \onlinecite{dierk}.

\begin{figure}
\centering{\resizebox*{!}{7.5cm}{\includegraphics{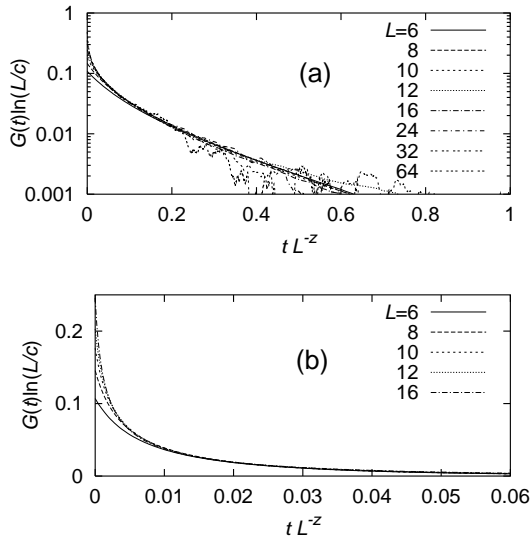}}}
\caption{Scaling of the supercurrent correlation function $G(t)$ in 2D at the KT transition ($T=0.9$) in case of the PBC for (a) RSJD and
(b) RD. The scaling collapse of the data shown is for $z=1.5$ and 2.0 in case of (a) RSJD and 
(b) RD, respectively. [The apparent spread of the data in (a) is only due to the 
logarithmic scale and insufficient convergence for the largest lattice sizes at large $t$.]}
\label{fig:2d_Gt}
\end{figure}

In order to determine $z$ at the KT transition for the PBC we use the following 
finite-size scaling form Eq.~(\ref{Lscale_2D}) (see Sec.~\ref{subsubsec:supercurrent})
of the supercurrent correlation function $G(t)$ 
\[ 
\ln \left(\frac{L}{c}\right)G(t)=F_G(tL^{-z}).
\]
Figure~\ref{fig:2d_Gt} shows the corresponding scaling plot at $T=0.90 \approx T_c$ 
both for RSJD and RD [Figs.~\ref{fig:2d_Gt}(a) and (b), respectively]. 
Very good scaling collapses are obtained in both cases with $z=1.5$ for RSJD
and $z=2.0$ for RD.
This clearly demonstrates that the value of $z$ for RSJD with the PBC is different from the 
expected value of $2$ which was obtained with the FTBC. In these scaling collapses one should 
note that the relaxation is much faster for RD than for RSJD, as is apparent by 
comparing the scales on the horizontal axes [note that vertical axis is in a 
logarithmic scale in 
Fig.~\ref{fig:2d_Gt}(a) and in a linear scale in Fig.~\ref{fig:2d_Gt}(b)]. 
It is also interesting to note that the value of 
the constant $c$ in the logarithmic correction of Eq.~(\ref{Lscale_2D}) comes from the 
static properties, as described in Appendix~\ref{append:scale}, 
and consequently should be independent of the dynamics. 
In accordance with this expectation the good scalings in Fig.~\ref{fig:2d_Gt}
are achieved with the same value of $c$: Both for RSJD in Fig.~\ref{fig:2d_Gt}(a) 
and for RD in Fig.~\ref{fig:2d_Gt}(b), we found that $c=0.60$ gives a good collapse.

\begin{figure}
\centering{\resizebox*{!}{7.5cm}{\includegraphics{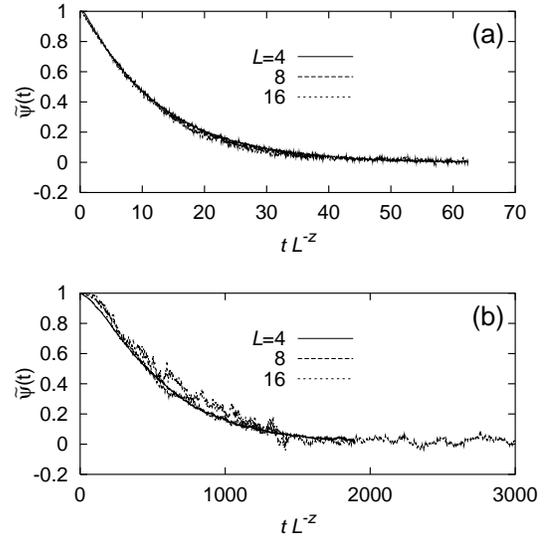}}}
\caption{Short-time relaxation for 2D RSJD at $T=0.9$ . (a) For the FTBC $z \approx 2.0$
and (b) for the PBC $z \approx 1.2$ are obtained. 
Note the enormous time scale for the PBC.
}
\label{fig:2drsj_short_Tc}
\end{figure}

\begin{figure}
\centering{\resizebox*{!}{7.5cm}{\includegraphics{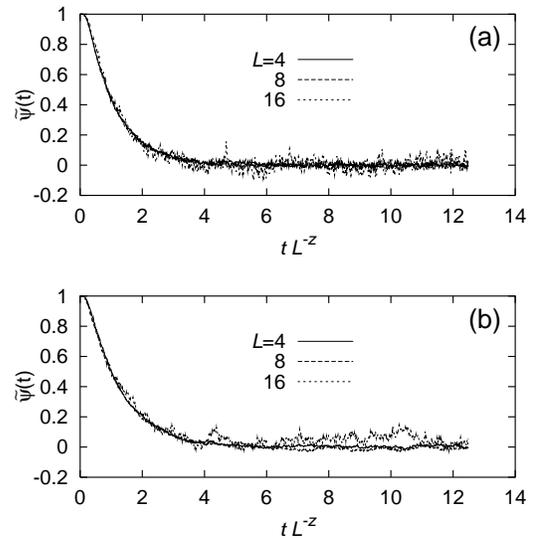}}}
\caption{Short-time relaxation for 2D RD at $T=0.9$  with the (a) FTBC and (b) PBC. 
For both boundary conditions $z \approx 2.0$ is obtained from the data collapse. Note that the decays for both boundary conditions have the same time scale.}
\label{fig:2dr_short_Tc}
\end{figure}

In Fig.~\ref{fig:2drsj_short_Tc}, we next show the decay of $\tilde\psi$
(see Sec.~\ref{subsec:shorttime} for details)
at $T=0.90 \approx T_c$ for RSJD with the (a) FTBC and (b) PBC, which demonstrates that 
$z \approx 2.0$ (for the FTBC) and $z\approx 1.2$ (for the PBC) result in good data 
collapses to scaling curves. However,
only the FTBC leads to the expected value $z \approx 2.0$. One should also note that
the PBC results in an extremely slow decay of $\tilde\psi$. Similarly
Fig.~\ref{fig:2dr_short_Tc} shows the 
decay of $\tilde\psi$ for RD at $T = 0.90$
with the (a) FTBC and (b) PBC. In both cases good data collapses are obtained for  
$z \approx 2.0$. In this case of RD,
both boundary conditions have the same magnitude of the decay time scale.
A possible interpretation is discussed in Sec.~\ref{sec:conc}.

\subsubsection{Low-temperature phase}
\begin{figure}
\centering{\resizebox*{!}{7.5cm}{\includegraphics{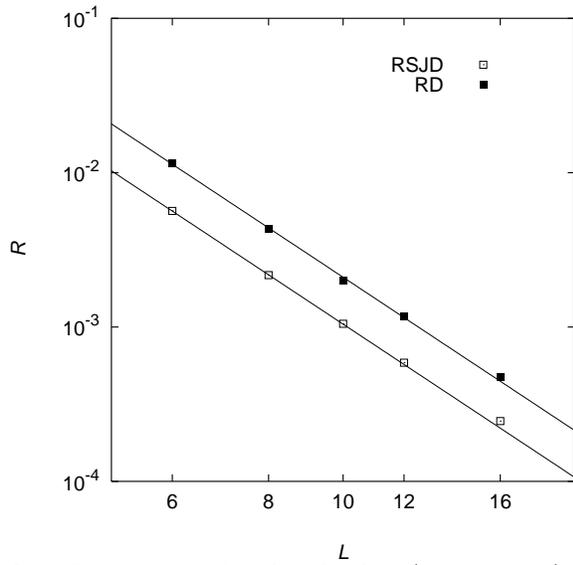}}}
\caption{Resistance scaling for 2D RSJD (open squares) and RD (solid squares) 
with the FTBC at $T$=0.8.  For both dynamics, $z \approx 3.3$ is obtained from the slopes of 
the lines in the figure. (The data points were taken from Ref.~\protect\onlinecite{bjkim99a}.)}
\label{fig:2d_R_low}
\end{figure}

The 2D $XY$ model is special in that the whole low-temperature phase is ``quasi'' critical. 
This means that each temperature in the low-temperature phase is characterized by a 
temperature-dependent dynamic critical exponent $z$.
Just as in the previous section, this temperature-dependent $z$ can
be determined from the size scaling of the linear resistance, i.e., $R\propto L^{-z}$.
Figure~\ref{fig:2d_R_low} shows the finite-size scaling
of resistance at $T = 0.8 (< T_c)$ for 2D RSJD and RD
with the FTBC (all data are from Ref.~\onlinecite{bjkim99a}) 
and we find $z \approx 3.3$ for both types of dynamics.
In Ref.~\onlinecite{bjkim99a}, this value has been compared with
$z_{\rm scale}$ and $z_{\rm AHNS}$ at this temperature
and it has been concluded that the observed value $3.3$
is very close to $z_{\rm scale}\approx 3.4$.

\begin{figure}
\centering{\resizebox*{!}{7.5cm}{\includegraphics{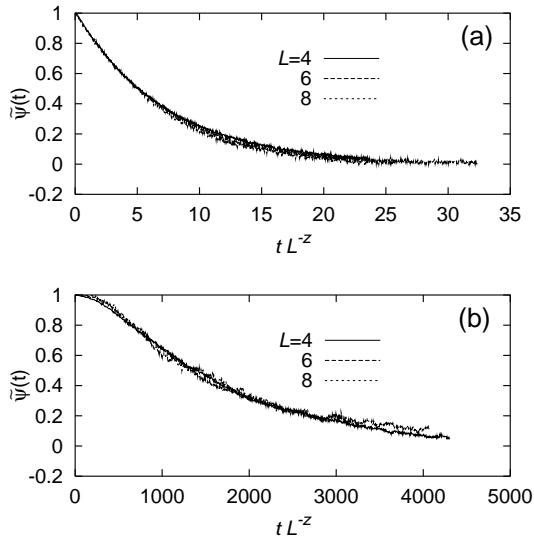}}}
\caption{Short-time relaxation for 2D RSJD at $T=0.80$. (a) For the FTBC $z \approx 3.2$
and (b) for the PBC $z \approx 1.4$ are obtained from the best data collapse. 
}
\label{fig:2drsj_short_low}
\end{figure}

\begin{figure}
\centering{\resizebox*{!}{7.5cm}{\includegraphics{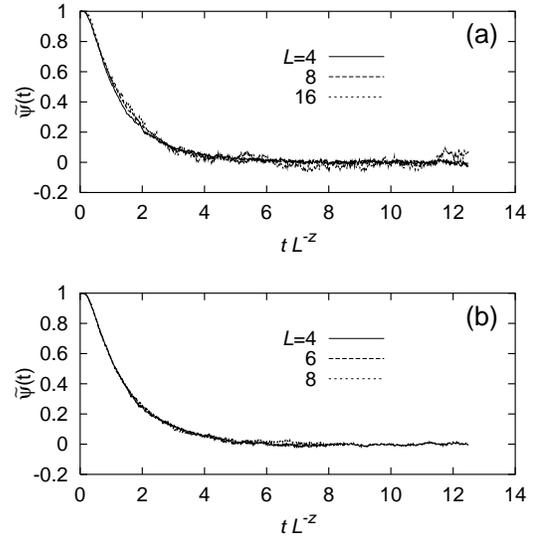}}}
\caption{Short-time relaxation for 2D RD at $T=0.80$ for the (a) FTBC and (b) PBC.
For both boundary conditions, $z\approx 2.0$ is found from the best data collapse.
}
\label{fig:2dr_short_low}
\end{figure}

In the same way as in the previous section the temperature-dependent $z$ in 
the low-temperature phase can also be probed by the short-time relaxation method
described in Sec.~\ref{subsec:shorttime}.
The divergence of the correlation length in the whole low-temperature phase in 2D
turns the finite-size scaling form~(\ref{eq:short}) into the simpler form 
\begin{equation}
\tilde\psi = F_\psi(t/L^z), 
\end{equation}
with the temperature-dependent $z$.
Figures~\ref{fig:2drsj_short_low} and \ref{fig:2dr_short_low}
show the finite-size scaling of the short-time relaxation 
at $T=0.80$ with the (a) FTBC and (b) PBC. The value $z \approx 3.2$
found in Fig.~\ref{fig:2drsj_short_low}(a) for RSJD with the FTBC
is in agreement with $z \approx 3.3$ obtained from the resistance scaling
in Fig.~\ref{fig:2d_R_low} within numerical accuracy.
We interpret this as an evidence that $z$ for the RSJD with the FTBC is 
indeed given by $z_{\rm scale}$. 
This is in contrast to the RSJD with the PBC for which at the same temperature ($T=0.80$)
$z\approx 1.4$ is determined from short-time relaxation as 
shown in Fig.~\ref{fig:2drsj_short_low}(b). 
As will be discussed in Sec.~\ref{sec:conc}, we interpret this as further evidence that,
in case of the 2D RSJD, $z$ does depend on the boundary condition. 
The short-time relaxation for RD gives a quite different result:
$z\approx 2$ is obtained for $T=0.80$  for both the FTBC and PBC, as shown in
Fig.~\ref{fig:2dr_short_low}.
If one compares this with the results at $T_c$ in Fig.~\ref{fig:2dr_short_Tc}, 
where $z\approx 2$ is also obtained, the implication is that the result 
$z\approx 2$ for the short-time relaxation is expected at any temperature 
in the low-temperature phase both for the FTBC and PBC. As will be discussed 
further in Sec.~\ref{sec:conc}, this suggests that the short-time relaxation for RD 
does not probe the true equilibrium critical dynamics.

\subsubsection{High-temperature phase}

\begin{figure}
\centering{\resizebox*{!}{7.5cm}{\includegraphics{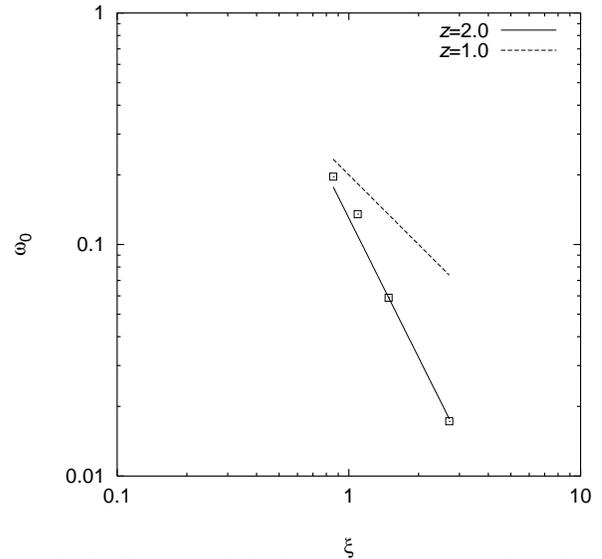}}}
\caption{Characteristic frequency $\omega_0$ determined
from the peak position of $|{\rm Im}1/\epsilon(\omega)|$ vs the
correlation length $\xi$
for 2D RSJD with the PBC at temperatures $T=1.0, 1.1, 1.2$, and 1.3
(from right to left).
The dynamic critical exponent $z$ defined by $\omega_0 \sim \xi^{-z}$
is shown to have a value close to $2.0$ (solid line). For comparison,
we also plot the dotted line which corresponds to $z = 1.0$.
}
\label{fig:2drsj_high}
\end{figure}

In the high-temperature phase there is a finite screening length $\xi$
which diverges as $T_c$ is approached from above. Close to $T_c$ one
then expects that the characteristic time scales as 
\[
\tau \sim \xi^z . 
\]
In case of the PBC, we can estimate $\xi$ and $\tau$ following the method
in Ref.~\onlinecite{jonsson}:  $\xi$ is obtained from the
wavevector dependence of the static dielectric function $1/\epsilon(0)$ introduced in
Eq.~(\ref{eq:epsilon0}). The characteristic frequency $\omega_0\sim 1/\tau$
is determined from the frequency dependence of $1/\epsilon(\omega )$;
$\omega_0$ is the position of the dissipation peak in $|{\rm Im}1/\epsilon(\omega)|$.
The result for RSJD with the PBC is shown in Fig.~\ref{fig:2drsj_high}, where
$z\approx 2$ is found from $\omega_0 \sim \xi^{-z}$. 
It should be noted that since this result is obtained for a temperature range where
$\xi/L \ll 1$ it is expected to be independent of the treatment of the boundary and 
hence applies to both the PBC and FTBC.
The same method applied to RD also gives $z \approx 2.0$ as shown in 
Ref.~\onlinecite{jonsson}.
In Fig.~\ref{fig:2drsj_high} the dotted line with slope $-1$ is also shown
and corresponds to the result in Ref.~\onlinecite{tiesinga}, where
$z\approx 1$ was obtained for RSJD with the PBC in the same temperature range.
Consequently, Fig.~\ref{fig:2drsj_high} implies, in contrast to Ref.~\onlinecite{tiesinga},
that $z = 2$ is the correct value for the PBC as well as for the FTBC,
when $z$ is determined from $\tau \sim \xi^z$.

\subsection{3D $XY$ model}
Next we turn to the 3D $XY$ model with current conserving RSJD and nonconserving  
RD, respectively. Both dynamic models have been used 
to describe the dynamic properties of high $T_c$ superconductors.
Whereas it is generally agreed upon that the 
static critical properties are those of the 3D $XY$ model in a region close to 
$T_c$ with the corresponding static critical exponents,~\cite{pasler98} there 
is less consensus on the dynamic critical properties. Several
seemingly mutually 
inconsistent experimental~\cite{pomar-holm,booth-jtkim,moloni97,jtkim99b} 
and simulational~\cite{weber97,ryu98,lidmar98} results have 
been reported. Similarly to  the 2D case described above in Sec.~\ref{subsec:result:2d}, 
we will here
arrive at a somewhat entangled picture by comparing values of $z$
obtained from the scalings in equilibrium for the two types of
dynamics (RSJD and RD) with the two types of boundary conditions (the PBC and FTBC), 
as well as from the short-time relaxation method by observing the time evolution 
towards equilibrium when starting from a nonequilibrium configuration.    
For convenience, the results of the simulations for the 3D $XY$ model 
are summarized in Table~\ref{table:3d}. 

\subsubsection{Resistance scaling} \label{subsubsec:3d_R}
We start with the determination of $z$ for the FTBC using
the finite-size scaling of the linear resistance, which is calculated
from the equilibrium fluctuation of twist variable $\Delta$ 
[see Eq.~(\ref{eq:Rmu})] A shorter presentation of
these results has also been given in Ref.~\onlinecite{jensen99}.
In 3D the correlation length diverges as $\xi\sim
|T-T_c|^{-\nu}$, making the extended scaling form 
of Eq.~(\ref{rscale_full}), as well as the intersection method in Eq.~(\ref{inter})
applicable in addition to the relation $R\sim L^{-z}$ at $T_c$.   

\begin{figure}
{\centering \resizebox*{!}{7.5cm}{\includegraphics{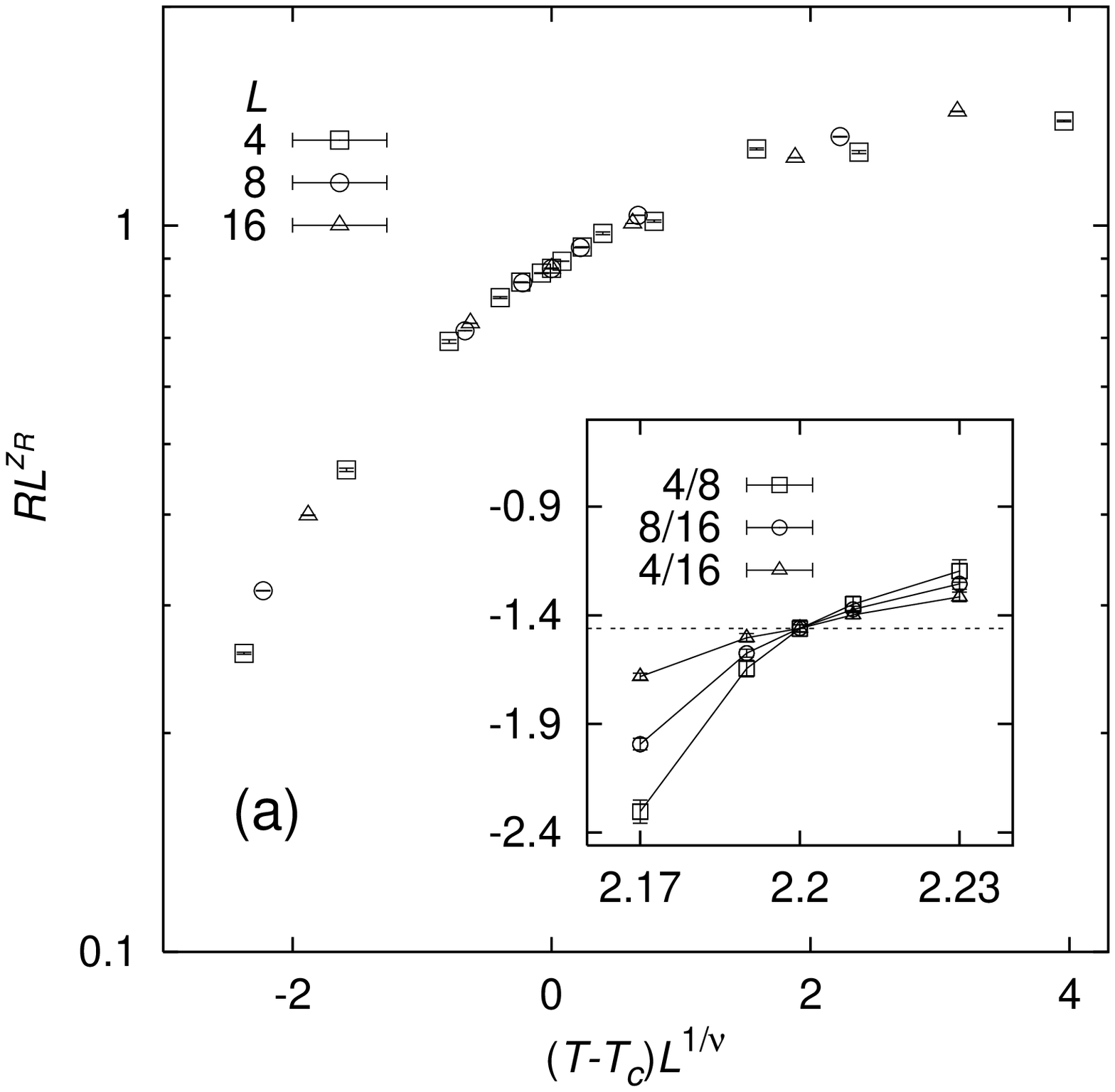}}\\
\resizebox*{!}{7.5cm}{\includegraphics{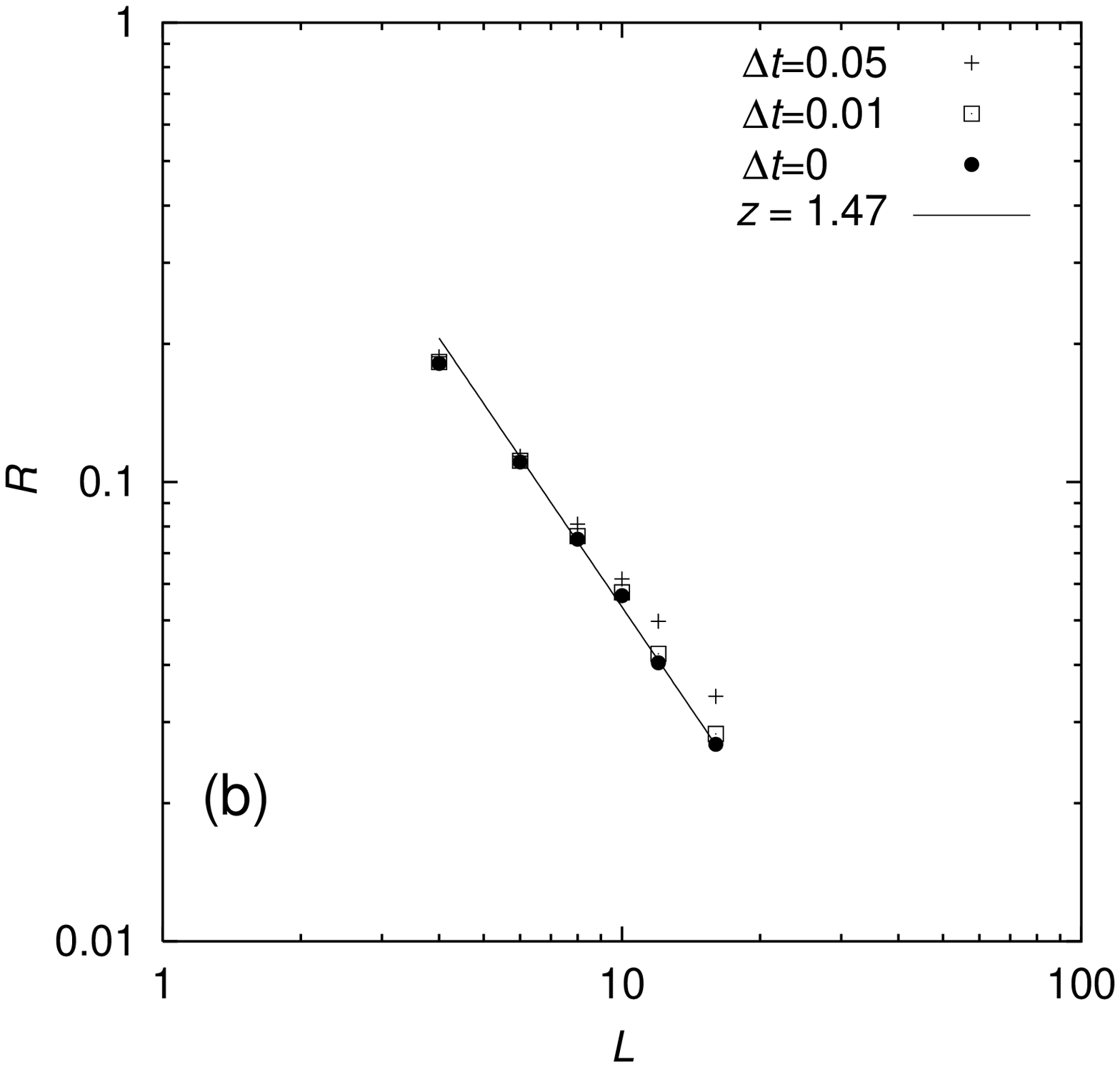}} \par}
\caption{(a) Scaling curve of the resistance $R$ for 3D RSJD with the FTBC
for $L=4$, 8, and 16 at $T=2.17$, 2.19, 2.20, 2.21, and 2.23
with parameter values $z = 1.46$ and $T_c = 2.200$ (both from 
the intersection method described in the text and shown in the inset), 
and the known value $\nu = 0.67$ (Ref.~\protect\onlinecite{olsson:Tc}).
(b) Determination of $z$ for 3D RD with the FTBC from 
the resistance scaling form $R\propto L^{-z}$ at $T_c=2.20$. 
The data points are for $L=4$, $8$, $10$, $12$, and $16$, and two integration time
steps $\Delta t=0.05$ and $0.01$. Linear extrapolation  to
$\Delta t=0$ gives $z=1.47$ from the least-squares fit.
}
\label{fig:3d_R}
\end{figure}

We first present the result for the scaling of the linear resistance for RSJD with the FTBC.
By using the intersection method explained in Sec.~\ref{subsubsec:resistance}
[see Eq.~(\ref{inter})] we determine $T_c$ and $z$ simultaneously from the unique
intersection point, as shown in the inset of Fig.~\ref{fig:3d_R}, which yields
$T_c \approx 2.200$ and $z \approx 1.46$. We then display in 
the main part of Fig.~\ref{fig:3d_R} the scaling plot of the linear
resistance [see Eq.~(\ref{rscale_full})], $RL^z$ as a function of the
scaling variable $(T-T_c)L^{1/\nu}$, at $T=2.17$, 2.19, 2.20, 2.21, and 2.23
for $L=4$, 8, and 16. Here we used $z$ and $T_c$ found from the intersection method, 
i.e., $z=1.46$ and $T_c=2.200$, respectively, and the known value
of the static exponent $\nu \approx 0.67$.~\cite{olsson:Tc}
We also tried to vary the values of $T_c$, $z$, and $\nu$ in the scaling
plot and concluded that $z = 1.46\pm 0.06$ for the case of
3D RSJD with the FTBC. It is noteworthy that $T_c \approx 2.200$ from the
intersection method is very close to the known value
of $T_c$ [$\approx 2.2018$ (Ref.~\onlinecite{olsson:Tc})] from the MC simulation.

For RD with the FTBC we only focus on the scaling relation
$R\sim L^{-z}$ at $T_c$, since it is found that the resistance for the RD case is harder
to converge due to a sensitivity of the result on the discrete time step
in the numerical integration of dynamic equations  even when
using the second-order RKHG algorithm.~\cite{batrouni85}
In contrast, we did not observe any significant sensitivity to the
time step in RSJD and we fix $\Delta t = 0.05$ throughout the present
work for RSJD.
In order to overcome the problem in RD 
due to the finite-time step we obtain data for two different 
time steps ($\Delta t = 0.05, 0.01$) and linearly extrapolate
to $\Delta t = 0$, as shown in Fig.~\ref{fig:3d_R}(b) for $L=4$, 6, 8, 10, 
12, and 16.
The slope of the line in the log-log plot of $R$ versus $L$ in Fig.~\ref{fig:3d_R}(b)
gives $z\approx 1.5$ also for 3D RD with the FTBC. 

\subsubsection{Supercurrent scaling}
\begin{figure}
{\centering \resizebox*{!}{7cm}{\includegraphics{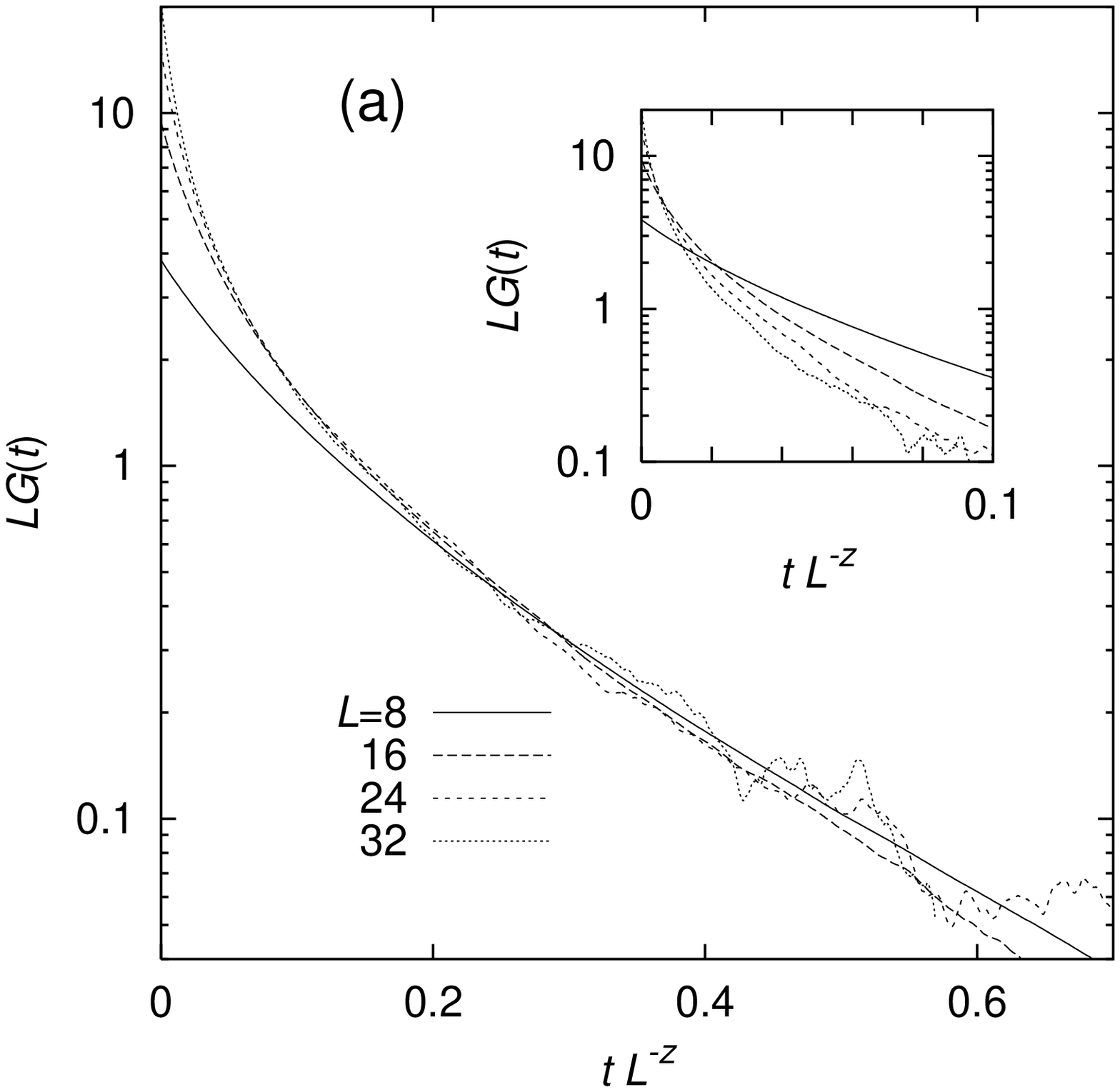}} \\
\resizebox*{!}{7cm}{\includegraphics{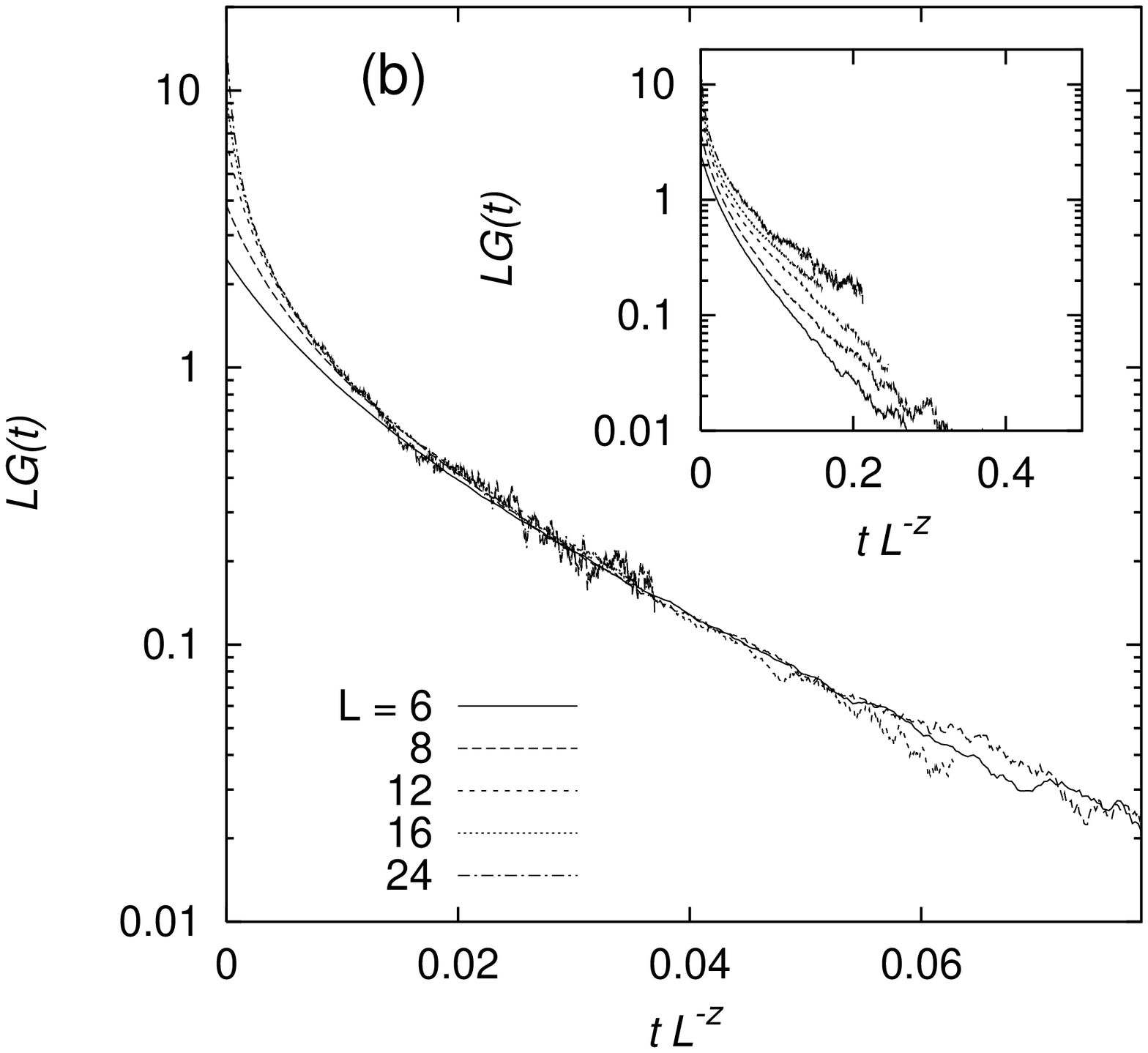}} \par}
\caption{Finite-size scaling at $T_c$ of the correlation
function \protect\( LG(t)\protect \) vs \protect\( t/L^{z}\protect \)
for 3D (a) RSJD with the PBC and (b) RD with the PBC. In the main parts of 
(a) and (b), $z=1.5$ and $z=2.05$ are shown to give good scaling collapses
for (a) RSJD and (b) RD, respectively, while in the insets the interchanged
values [$z=2.0$ and $z=1.5$ for (a) and (b), respectively] are shown to be 
inconsistent with the scaling collapse.}
\label{fig:3d_super_L}
\end{figure}

For the PBC we use the scaling of the supercurrent
correlation function $G(t)$ introduced in Sec.~\ref{subsubsec:supercurrent}
in order to obtain $z$.
In Fig.~\ref{fig:3d_super_L}(a) we use the finite-size scaling 
form in Eq.~(\ref{Lscale_3D}) and plot $LG$ as a function 
of the scaling variable $t/L^z$ for (a) $L=8$, 16, and 24 for RSJD
and (b) $L=6$, 8, 12, 16, and 24 for RD, respectively: 
Optimal data collapse is achieved for $z=1.5$ (RSJD), and $z=2.05$ (RD), respectively.
In the insets of Fig.~\ref{fig:3d_super_L} we use (a) $z=2.0$ for RSJD and (b) 
$z=1.5$ for RD, respectively, and show that the data collapse
becomes significantly worse and consequently conclude that the
$z$ values obtained by  this data collapse method 
are well determined [see the main parts of Figs.~\ref{fig:3d_super_L}(a) and 
\ref{fig:3d_super_L}(b)].
One notes that for RSJD $z\approx 1.5$ is obtained for both
the FTBC and PBC, whereas for RD $z\approx 1.5$ and $z \approx 2$ are
obtained for the FTBC and PBC, respectively.
    
\begin{figure}
{\centering \resizebox*{!}{7cm}{\includegraphics{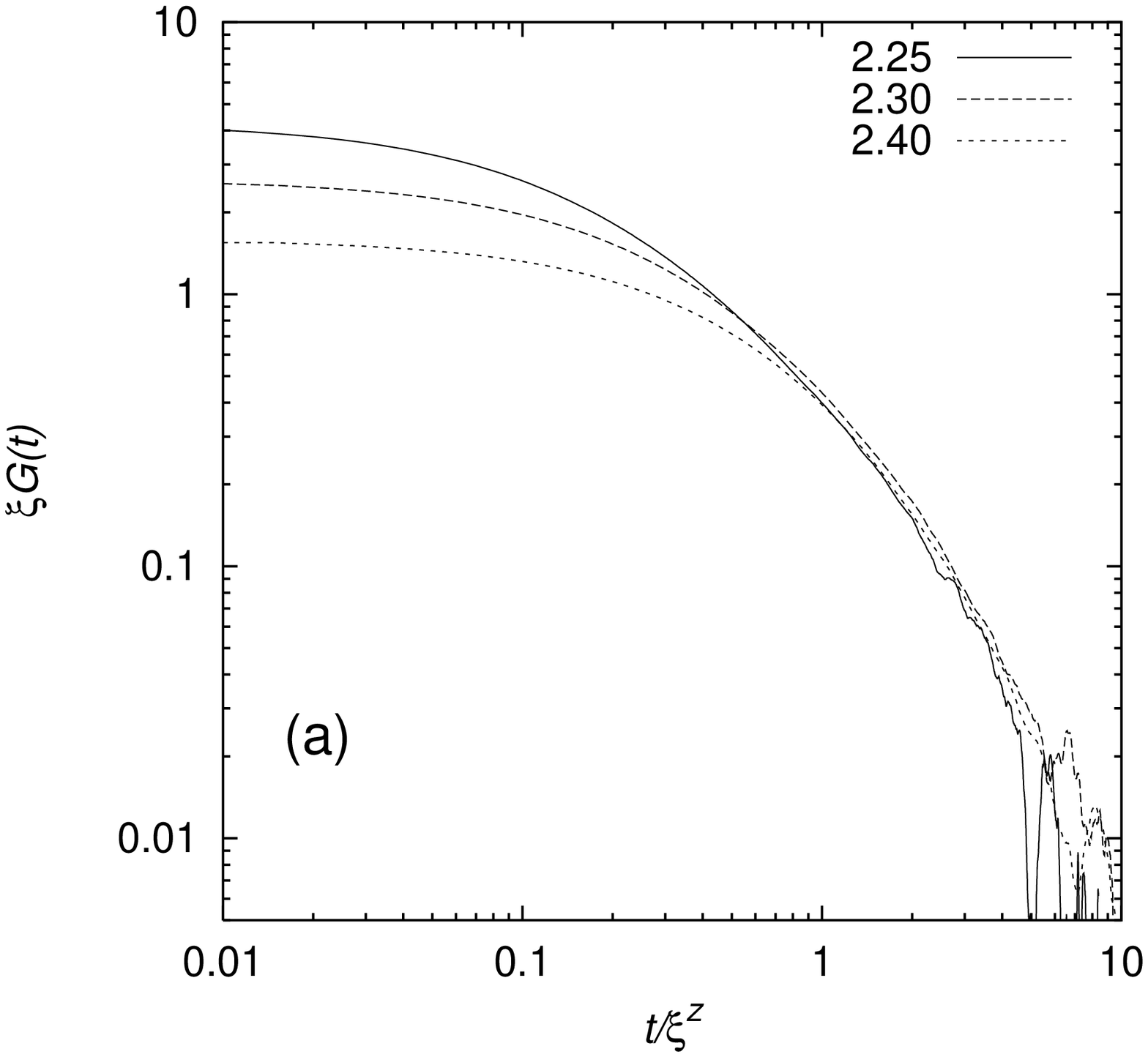}} \\
\resizebox*{!}{7cm}{\includegraphics{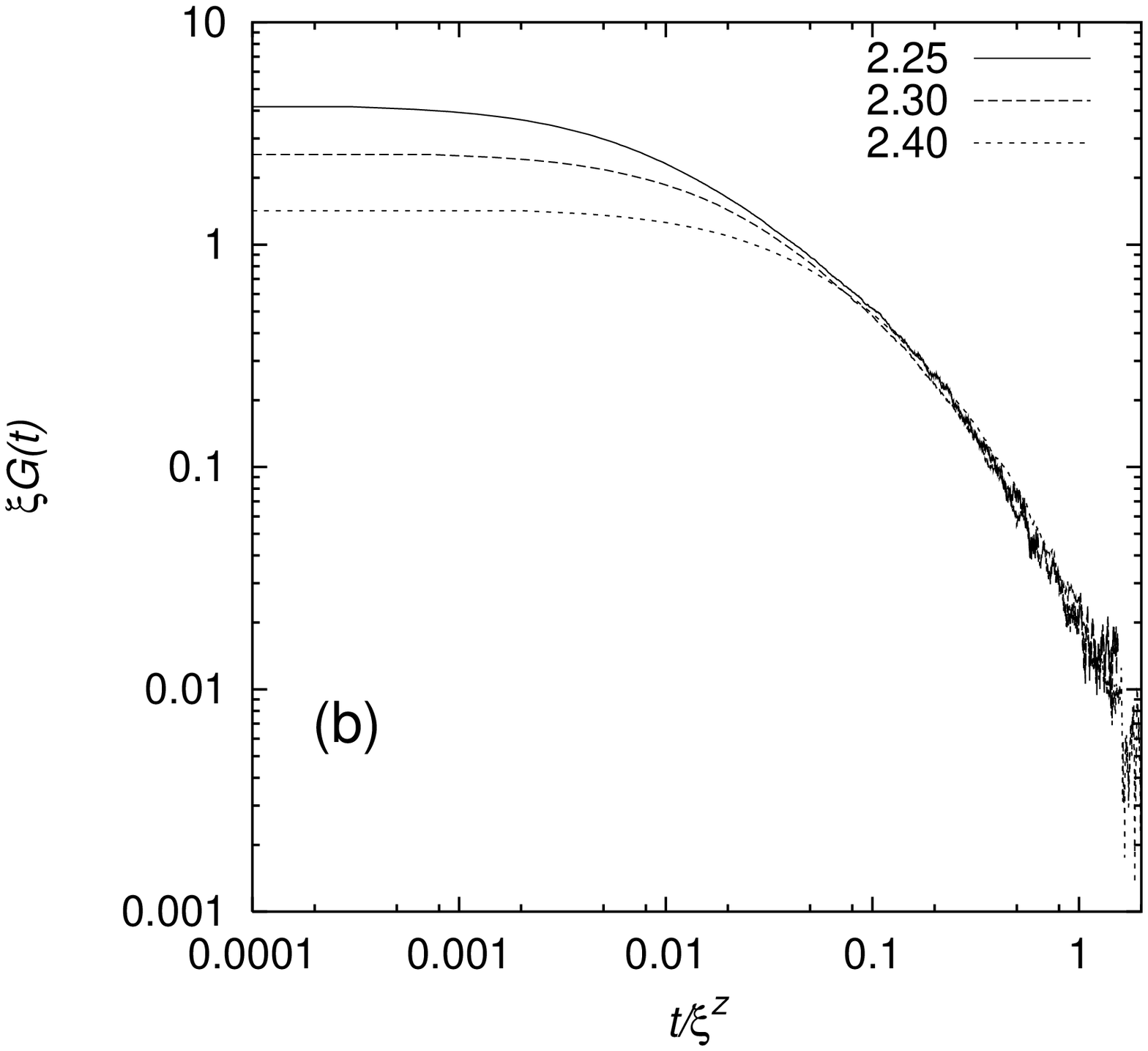}} \par}
\caption{Scaling of the correlation function $\xi G(t)$ above $T_c$
is shown against the scaling variable $t \xi^z$ for 3D
(a) RSJD with the PBC and (b) RD with the PBC for $L=24$ and $T=2.25$, 2.30, and 2.40.  
From the scaling collapse, with $\xi$ from the MC simulation (Ref.~\protect\onlinecite{olsson:3dxi}), 
we obtain (a) $z\approx 1.4$ for RSJD and (b) $z\approx 1.9$ for RD. 
}
\label{fig:3d_super_xi} 
\end{figure}

In the critical region above $T_c$ where $\xi \ll L$
we instead have $\xi G(t)$ as a scaling function with the scaling
variable $t/\xi^z$ [see Eq.~(\ref{xiscale})].
Figure~\ref{fig:3d_super_xi} shows this scaling results 
at temperatures above $T_c$ ($T=2.25$, 2.30, and 2.40) for $L=24$. 
By comparing with the results for $L=32$, it is explicitly checked 
that there remain no significant finite-size effects in the current
temperature range. 
In Fig.~\ref{fig:3d_super_xi} the corresponding values of $\xi$ are 
taken from high precision MC simulations.~\cite{olsson:3dxi}
As shown in Fig.~\ref{fig:3d_super_xi},
the optimal value $z=1.4(2)$ is found for RSJD and
$z=1.9(2)$ for RD, respectively, which is consistent with the
finite-size scaling of $G(t)$ at $T_c$. However, we note that the determination 
of $z$ in the case of the finite-$\xi$ scaling above $T_c$ yields a somewhat larger 
uncertainty. Furthermore, $z\approx 2$ found for RD with the PBC is particularly 
intriguing to understand since we expect that this result should be 
independent of boundary condition in this high-temperature regime 
where $\xi \ll L$: Thus one expects the same value $z \approx 2$ for the FTBC 
in this high-temperature regime. This in turn suggests a 
discontinuous jump in the $z$ value from $z \approx 2$ to $z \approx 1.5$
as $T_c$ is approached from above, since $z \approx 1.5$ at $T_c$ was observed
in the scaling of the linear resistance in Sec.~\ref{subsubsec:3d_R}.
This possibility is also discussed in Sec.~\ref{sec:conc}.

\subsubsection{Short-time relaxation scaling}
\begin{figure}
\centering \resizebox*{!}{7.5cm}{\includegraphics{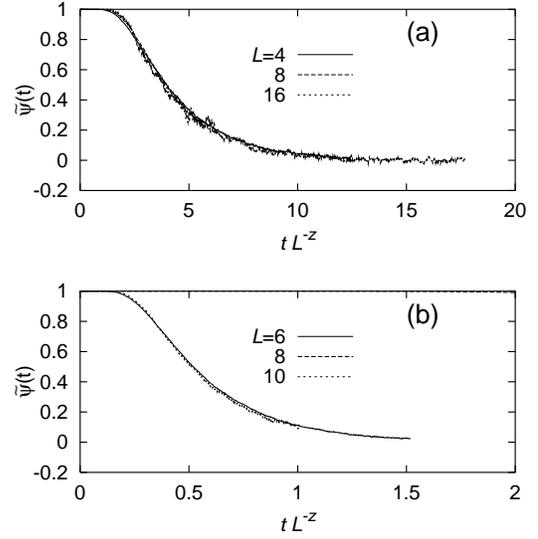}}
\caption{Short-time relaxations of $\tilde\psi$ in 3D with the FTBC at $T=T_c=2.20$ are shown 
as functions of the scaling variable $tL^{-z}$ for (a) RSJD with $L=4$, 6, and 16,
and for (b) RD with $L=6$, 8, and 10.
From the scaling collapse $z = 1.50$ and  $z = 1.95$ are found to yield
smoothly collapsed single curves for (a) RSJD and (b) RD, respectively.
}
\label{fig:3d_short_ftbcTc}
\end{figure}

\begin{figure}
{\centering \resizebox*{!}{7cm}{\includegraphics{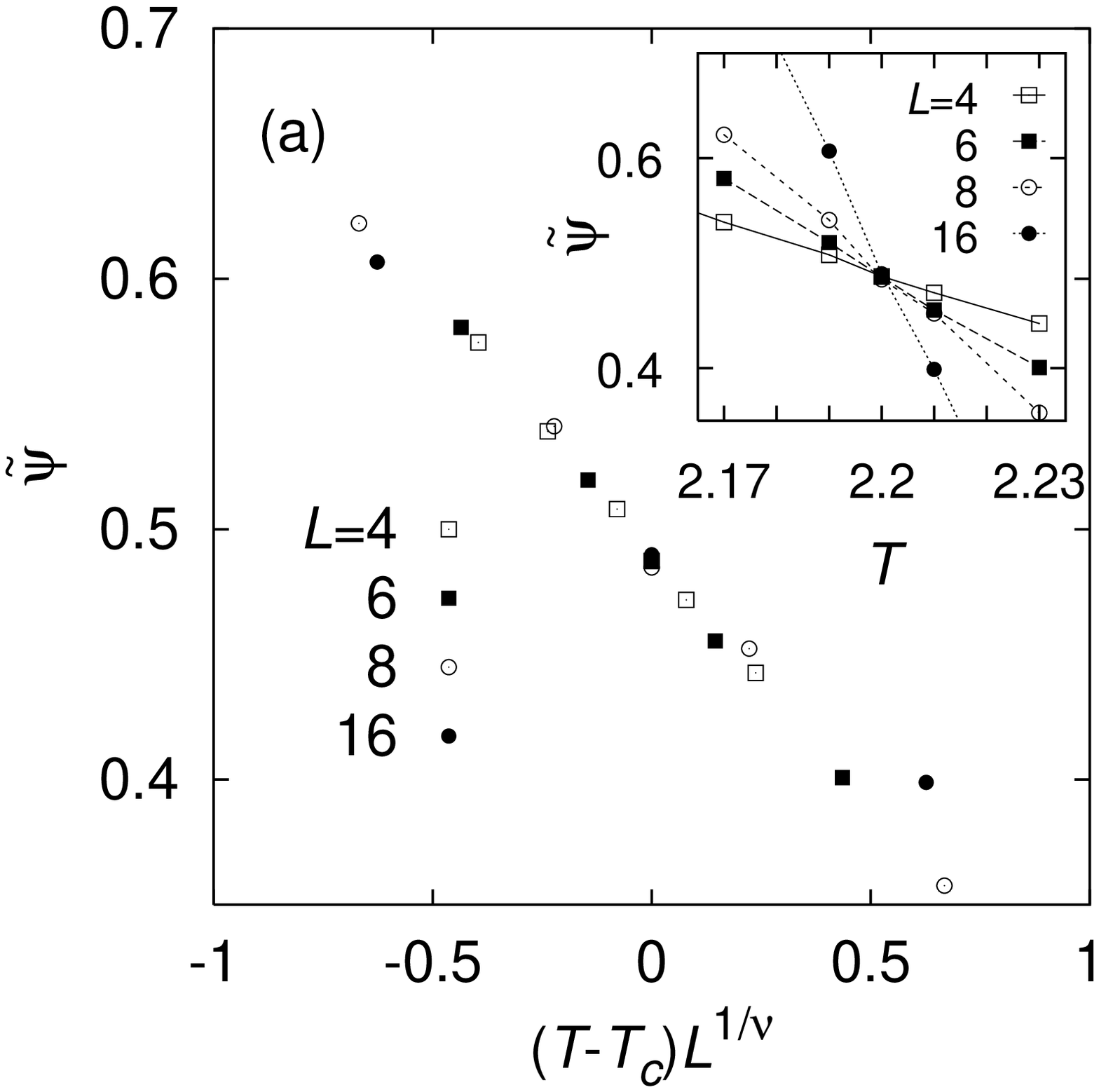}}\\
\resizebox*{!}{7cm}{\includegraphics{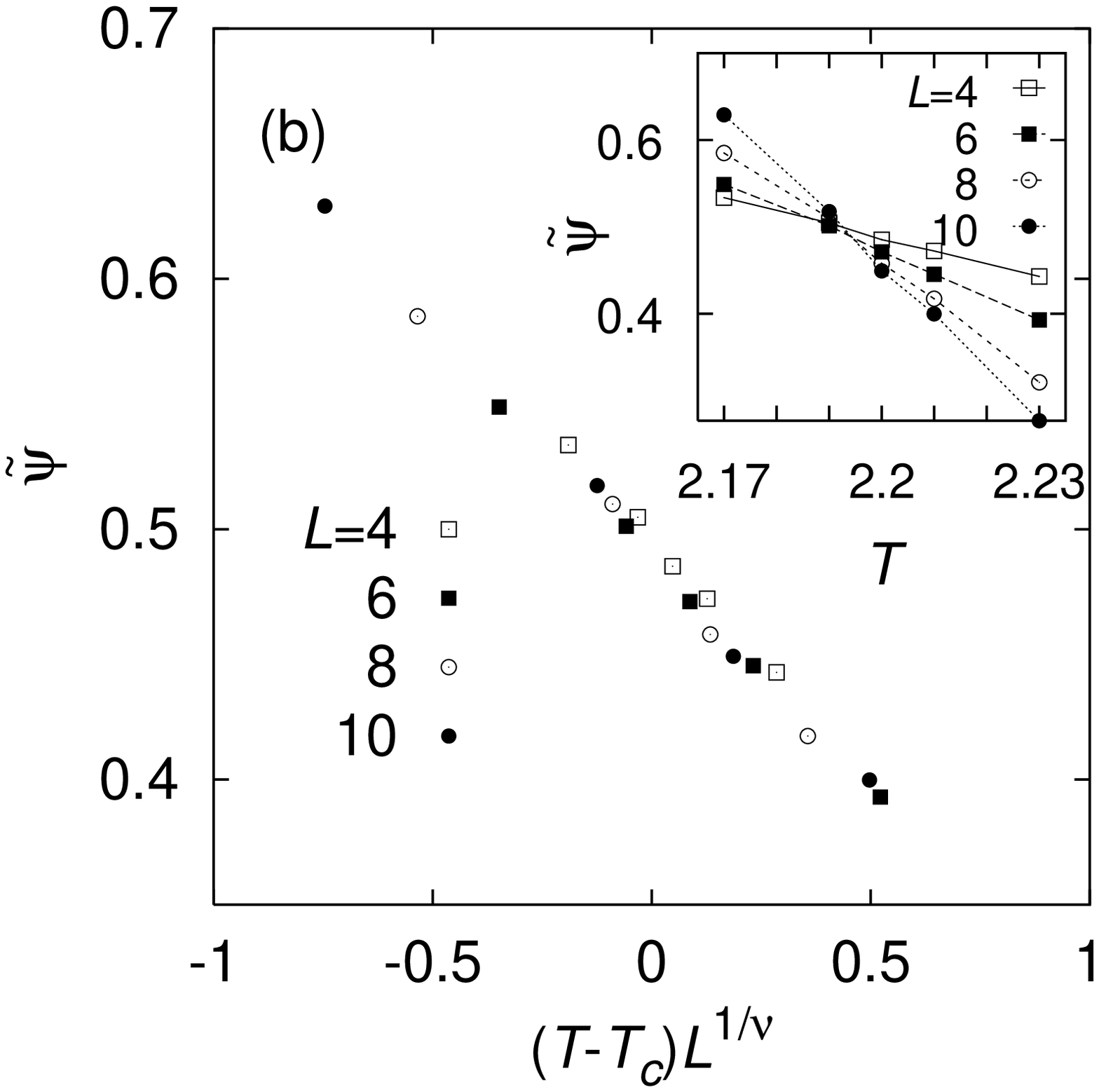}} \par}
\caption{Finite-size scaling of the short-time relaxation of $\tilde\psi$ for
3D (a) RSJD for $L=4$, 6, 8, and 16 and (b) RD for $L=4$, 6, 8, and 10
both with the FTBC and at $T=2.17$, 2.19, 2.20, 2.21, and 2.23.  
As shown in insets, the scaling function 
$\tilde{\psi}=F_\psi\bigl(t/L^z,(T-T_c)L^{1/\nu}\bigr)$ with a fixed $a=t/L^z$ 
suggests the existence of a single crossing point (a) at $T_c = 2.200$
with $(z,a)=(1.5,4.0)$ for RSJD and (b) at $T_c = 2.194$ with $(z,a)=(2.0,0.5)$ 
for RD. All data points in the insets collapse onto a single smooth curve 
with the scaling variable $(T-T_c)L^{1/\nu}$ for both models with the known 
value $\nu = 0.67$ (Ref.~\protect\cite{olsson:Tc}), as shown in main parts.}
\label{fig:3d_short_ftbc}
\end{figure}

The short-time relaxation method described in Sec.~\ref{subsec:shorttime}
probes the relaxation towards equilibrium from a nonequilibrium configuration. 
We start with the presentation of the results obtained for RSJD and RD
with the FTBC.  Using the scaling form of $\tilde\psi$ in Eq.~(\ref{eq:short}) at $T_c$, 
where the scaling function has only one scaling variable $t/L^z$, 
we first show in Fig.~\ref{fig:3d_short_ftbcTc} the scaling plot of $\tilde\psi$ 
at $T = 2.20$ for (a) RSJD with $L=4$, 8,  and 16 and (b) RD with $L=6$,
8, and 10, respectively. All the data can be made to collapse onto a single curve in a broad
range of the scaling variable for $z \approx 1.5$ and $z\approx 2.0$ for
RSJD and RD, respectively. However, the above method presumes {\it a priori} knowledge 
of $T_c$. To circumvent this, one can alternatively use an intersection 
method with a fixed value of $a = tL^{-z}$ in the first argument of the scaling 
form in Eq.~(\ref{eq:short_a}) (see Sec.~\ref{subsec:shorttime}).
In insets of Fig.~\ref{fig:3d_short_ftbc} we display data points at $T=2.17$, 2.19, 2.20, 2.21, 
and 2.23 for (a) RSJD with $L=4$, 6, 8, and 16 and (b) RD with $L=4$, 6, 8, and 10,
and show the results of the iterative intersection method. We obtain again 
$z\approx 1.5$ and $z\approx 2.0$, as well as the estimations of the critical temperatures 
$T_c \approx 2.200$ and $T_c \approx 2.194$ for (a) RSJD and (b) RD, respectively.
We believe that the existence of an unique intersection point in each dynamics
with the value $T_c \approx 2.200$ obtained for RSJD, which is in very good agreement
with  $T_c \approx 2.200$ obtained previously from the resistance scaling, 
and with $T_c \approx 2.2018$ from MC simulations,~\cite{olsson:Tc}
make this short-time relaxation method very reliable.
One notes that the slight temperature shift for RD is again the effect of
the finite time step, as already observed in the calculation of the linear resistance.
We have also checked the dependence on $a$ values and observed no significant changes
in resulting values of $T_c$ and $z$ in a broad range where $0.4 < \tilde\psi < 0.9$.
Using $z$ and $T_c$ found from the intersection method, we in Fig.~\ref{fig:3d_short_ftbc}
confirm that the full scaling form is borne out to high precision
with $\nu=0.67$ determined from MC simulations.~\cite{olsson:Tc}

\begin{figure}
\resizebox*{!}{6.0cm}{\includegraphics{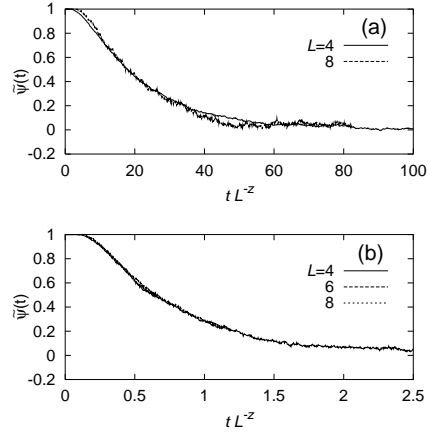}}
\caption{Short-time relaxation of $\tilde\psi$ in 3D with the PBC at $T=T_c = 2.20$ 
as functions of the scaling variable $tL^{-z}$ for (a) RSJD with $L=4$ and 8, 
and for (b) RD with $L=4$, 6, and 8.
From the scaling collapse $z \approx 1.2$ and  $z \approx 2.1$ are found 
for (a) RSJD and (b) RD, respectively.}
\label{fig:3d_short_pbc}
\end{figure}

We next consider the short-time relaxation for PBC, and show in
Fig.~\ref{fig:3d_short_pbc} the scaling plot at $T=2.20$, 
$\tilde{\psi}=F_{\psi}(t/L^z)$, for (a) RSJD and (b) RD.
Treating $z$ as a free parameter, we obtain $z\approx 1.2$ and $z\approx 2$
for RSJD and RD, respectively. This suggests that the value for RSJD 
with the PBC is lower than $z\approx 1.5$ obtained from the same short-time
relaxation method for RSJD with the FTBC,  whereas for RD a value $z\approx 2$ is obtained
both for the FTBC and PBC. 
As already observed in 2D, RSJD with the PBC has a  very large decay time scale 
in contrast to RSJD with the FTBC  as well as to RD with both the PBC and FTBC.

\subsection{4D $XY$ model}
\begin{figure}
{\centering \resizebox*{!}{6.5cm}{\includegraphics{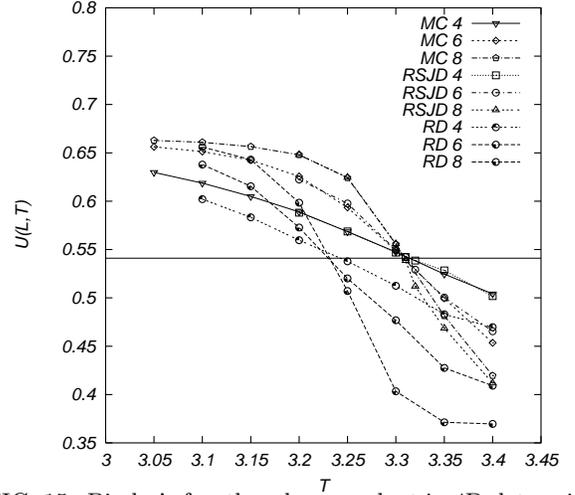}} \\
\par}
\caption{Binder's fourth order cumulant in 4D determined from MC simulations, 
RSJD, and RD for $L=4$, 6, and 8 at several temperatures
between $T= 3.05$ and $T=3.40$. Data points for MC and RSJD coincide within error
bars and give $T_c\approx3.31$, whereas the
RD results show a relative large dynamic shift to $T_c\approx3.25$.}
\label{fig:4d_binder}
\end{figure}

For completeness we also determine $z$ in 4D.~\cite{jensen-lt22}
As a prerequisite we first estimate $T_c$ through the use of 
MC simulations in conjunction with the finite-size 
scaling analysis of the Binder's fourth-order cumulant~\cite{binder} $U$,
which is independent of $L$ precisely at $T_c$, 
\[ 
U(L,T)=1-\frac{\langle |m|^4 \rangle}{3\langle|m|^2\rangle^2} , 
\] 
with the order parameter $m=\sum_{\bf r}e^{i\theta_{\bf r}}/L^4$. 
The results are shown in Fig.~\ref{fig:4d_binder} and $T_c \approx 3.31$ is found
from MC simulation with the PBC, which is consistent with earlier 
reports~\cite{bukenov93} but has a higher accuracy.
From the MC simulations we also verified that $\nu$ in 4D has
the expected mean-field value $\nu=1/2$ (see, e.g.,  Ref.~\onlinecite{goldenfeld}).
Since, as noted in the previous section, the size of the discrete time step in 
the integration of the dynamic equations of motion can lead to an effective 
increase of temperature, we explicitly determine the effective $T_c$ for RSJD and RD
with the time step $\Delta t=0.05$ from the crossing point of $U(L,T)$:
Figure~\ref{fig:4d_binder} shows that there is no significant difference 
between the effective and nominal temperature for RSJD, leading to 
$T_c({\rm RSJD}) \approx T_c({\rm MC}) \approx 3.31$. 
On the other hand, for RD we from the crossing point
obtain $T_c\approx 3.25$ at the same time step $\Delta t=0.05$, in
parallel with what was found for RD in 3D. 
It is to be noted that although the above critical temperatures have been
obtained only with the PBC, the same critical temperature is expected also for
the FTBC since all static quantities such as $T_c$ should not depend
on boundary conditions used. 

\begin{figure}
{\centering \resizebox*{!}{7cm}{\includegraphics{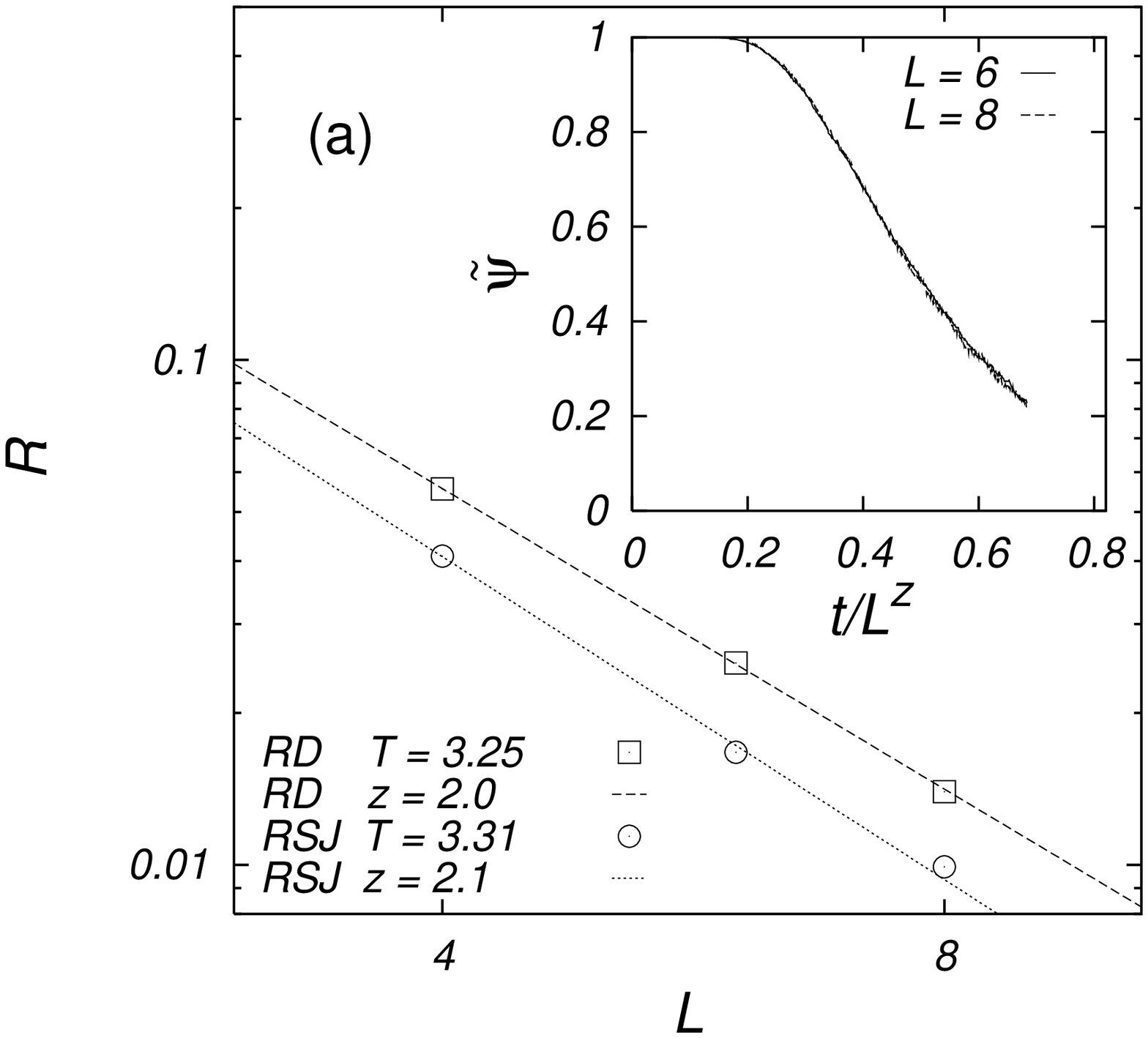}} \\
\resizebox*{!}{7cm}{\includegraphics{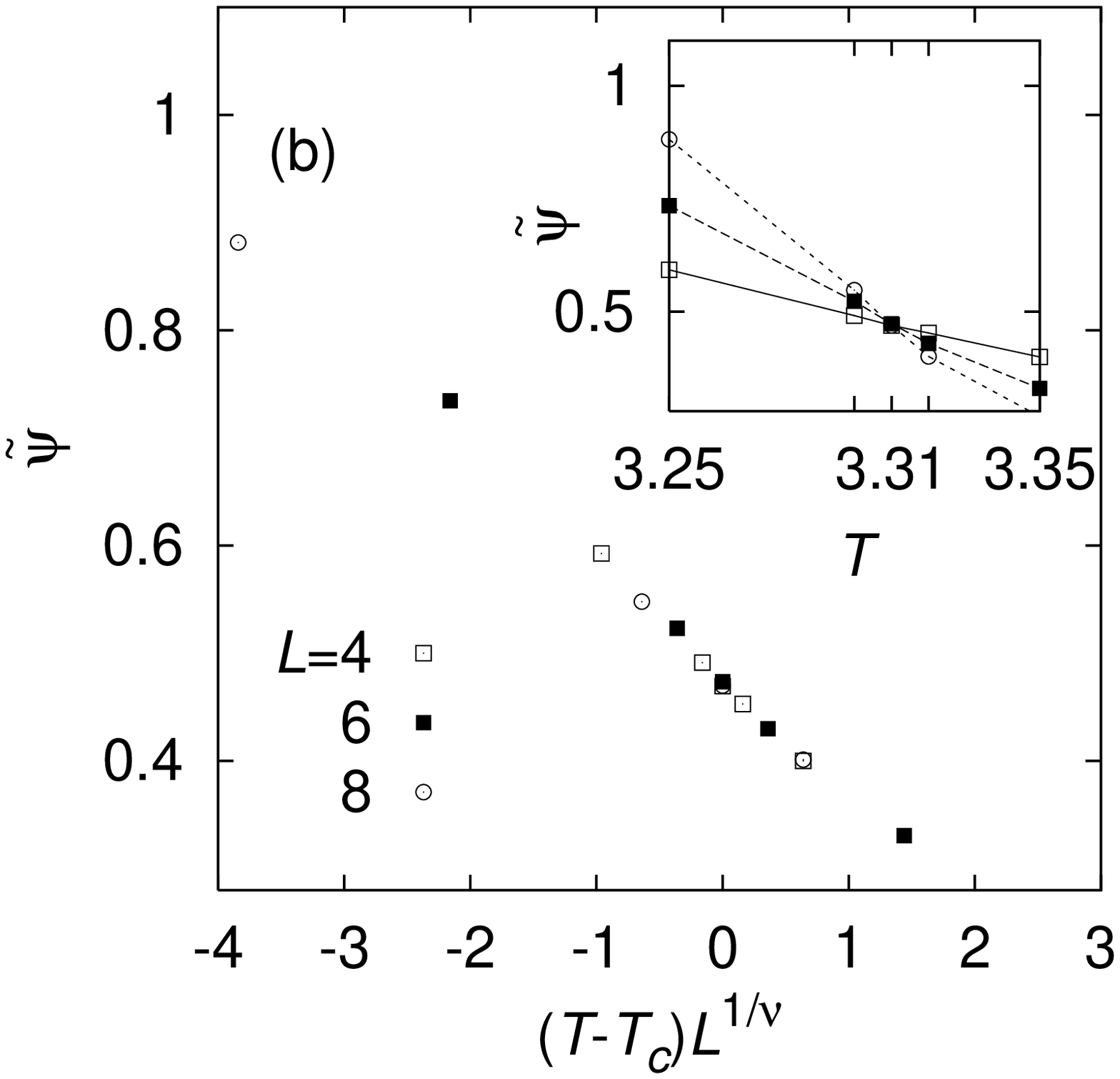}} \par }
\caption{(a) Determination of $z$ in 4D $XY$ model with the FTBC
from the resistance scaling form $R \sim L^{-z}$ for $L=4$, 6, 
and 8 at $T_c = 3.31$ for RSJD and at $T_c = 3.25$ for RD 
(see Fig.~\protect\ref{fig:4d_binder}).
From the slopes in log-log plot, $z \approx 2$ is concluded for
both dynamics. Inset: Short-time relaxation of $\tilde\psi$ for 4D
RD with the FTBC at $T=T_c=3.25$ is shown against the scaling variable
$tL^{-z}$ for $L=6$ and 8 and $z=2.0$ is found.
(b) Finite-size scaling of $\tilde\psi$ for 4D RSJD with the FTBC
for $L=4$, 6, and 8 at $T=3.25$, 3.30, 3.31, 3.32, and 3.35.
As shown in inset, the intersection method gives $T_c \approx 3.31$
with $z=2.0$. The main part displays the full scaling plot
of the form $\tilde{\psi}=F_\psi\bigl(t/L^z,(T-T_c)L^{1/\nu}\bigr)$
with $a = tL^z = 2.8$ and the mean-field value $\nu = 1/2$.
}
\label{fig:4d_R}
\end{figure}

Once $T_c$ is known from the calculation of the Binder's cumulant,
we can use the simple finite-size scaling form Eq.~(\ref{rscale})
for the linear resistance calculated with the FTBC by Eq.~(\ref{eq:Rmu})
(we use $\Theta = 2000$ for both RSJD and RD).
In Fig.~\ref{fig:4d_R}(a), we plot the linear resistance $R$ versus
$L$ at $T=3.31$ (RSJD) and $T=3.25$ (RD), and 
from the least-square fit we find $z\approx 2.1$ for RSJD 
and $z\approx 2$ for RD, respectively. 
In addition, we also measure the short-time relaxation with the FTBC
and present the result for RD at $L=6$ and 8
in the inset of Fig.~\ref{fig:4d_R}(a) by using the simple scaling 
form at $T = T_c = 3.25$, i.e., $\tilde\psi = F_\psi (tL^{-z})$,
which yields $z \approx 2.0$ in accordance with the result from the
resistance scaling. 
For RSJD with the FTBC, we construct the intersection plot 
for the short-time relaxation (similar to Fig.~\ref{fig:3d_short_ftbc} for 3D)
as displayed in the inset of Fig.~\ref{fig:4d_R}(b), and get
$z \approx 2.0$ and $T_c \approx 3.31$ from the unique crossing point.
It is interesting to note that the critical
temperature obtained here for RSJD with the FTBC is in a perfect
agreement with $T_c$ found from the Binder's cumulant method
for the other boundary condition, the PBC.
We then make the full scaling plot for $\tilde\psi$
in the main part of Fig.~\ref{fig:4d_R}(b) with the mean-field
value $\nu = 0.5$ and the estimated values $T_c = 3.31$ and $z=2.0$ above,
resulting in a very smooth collapse.

In short, we get $z\approx 2$ in 4D with the FTBC regardless of the dynamics 
we use (see Table~\ref{table:4d} for a summary of results); this is reassuring since the value $z=2$ is usually expected
in 4D where the phase transition acquires a mean-field nature.~\cite{hohenberg77,goldenfeld}

\section{Discussions and Comparisons}  \label{sec:conc}
As is clear from the simulation results presented in Sec.~\ref{sec:result}
for two-, three-, and four-dimensional $XY$ models, 
a very entangled picture emerges as regards to the dynamic critical exponent $z$.
In this section we discuss the main features.

\subsection{Discussions of the 2D $XY$ model} \label{subsec:dis2d}
We start our discussion with 2D (see Table~\ref{table:2d} for summary
of results) and first focus on RSJD at the KT transition. For a 2D
superfluid and superconductor the most widely expected value is $z=2$
although there have been a few different suggestions (Refs.~\onlinecite{tiesinga,pierson,dierk}).
The value $z=2$ can be inferred from the observed nonlinear current-voltage ($I$-$V$) exponent
$a=3$ (Refs.~\onlinecite{bjkim99a,PRL-z,kenneth}, and ~\onlinecite{weber96}) together with the
scaling argument that $a=z+1$.~\cite{fisher91,dorsey91} It may also be
directly obtained from the simple argument of the 
escape over the boundary presented in Sec.~\ref{subsec:result:2d} with the result
$z=1/\tilde{\epsilon}T^{CG}-2$, combined with the universal jump condition at
the KT transition $1/\tilde{\epsilon}T^{CG}=4$,~\cite{minnhagen:rev,nelson77,minnhagen81}
which leads to $z=2$ at the transition. For the 2D $XY$ model,
the KT transition temperature is $T_c\approx 0.9$ (Ref.~\onlinecite{olsson:Tc})
and as seen from Table~\ref{table:2d}, RSJD with the FTBC 
does give the expected value. However, RSJD with
the conventional PBC does in fact not give the expected
value: the supercurrent scaling gives $z\approx 1.5$ and the short-time
relaxation gives $z\approx 1.2$.

In order to understand the role played by the boundary conditions we 
consider a system with an open boundary, which is appropriate to describe
a superconducting film and a film of $^4$He in usual experiments. 
In such a case, when a
vortex-antivortex pair is introduced into the ground state
and then annihilated across the boundary, the system relaxes
back to the original ground state. 
The FTBC has been designed to keep the advantage of the PBC, which reduces
the finite-size effect compared to the free boundary condition,
as much as possible, while allowing this relaxing back. 
This relaxing back is,  however,  prohibited by the conventional PBC.~\cite{olsson} 
One may note that the escape-over-barrier 
argument in Sec.~\ref{subsec:result:2d} implicitly presumes this relaxation back
as a part of the escape process. 
One should also note that, when comparing to experiments with open boundaries, 
the FTBC has to be used in simulations instead of the PBC whenever the relaxation process 
across the boundary is important. This perspective suggests that the observed
difference between the FTBC and PBC at the KT transition for RSJD is 
due to the additional constraint on the physics caused by the PBC.

This can be substantiated somewhat further by studying
the low-temperature phase in 2D, where an ubiquitous ``quasi'' criticality 
with a diverging correlation length makes the critical finite-size scaling method
applicable. In Ref.~\onlinecite{bjkim99a}, $z\approx 3.3$
at $T=0.80$ was found for the FTBC from the resistance scaling in agreement
with the expected value $z=1/\tilde{\epsilon}T^{CG}-2 \approx 3.4$ within numerical errors. 
However, an estimate of the equilibrium scaling for the PBC at $T=0.85$ in Ref.~\onlinecite{bjkim99a} 
gave $z\approx 1.6$ instead of the expected result $z\approx 2.8$ (see Fig.~3 in
Ref.~\onlinecite{bjkim99a}). Thus in this low-temperature
phase $z$ determined with the PBC appears to be smaller ($z<2$) than the one with
the FTBC ($z>2$). However, the value for the FTBC is the relevant
one when comparing with experiments.

The situation above $T_c$ is as follows: The finite linear resistance $R$
calculated from the fluctuations of ${\bf \Delta}$ for the FTBC [see Eq.~(\ref{eq:Rmu})]
can be related to the conductivity calculated for the PBC through the connection
$R={\rm Re}\left[ 1/\sigma (\omega=0)\right]$ with $\sigma(\omega)$ in 
Eq.~(\ref{eq:sigma}).
We have explicitly checked this relation in our simulations at $T=1.4$,
by comparing the two values for the FTBC and PBC, respectively, 
and found good agreement.~\cite{check} 
From this observation, we expect that in this high-temperature phase 
where the correlation length is smaller than the linear size of the system, 
$R$ and $z$ are independent of boundary conditions.
Furthermore, in this temperature regime, transport properties like the linear
resistance are dominated by free vortices (with density $n_F$) 
and accordingly we expect $R\propto n_F \propto \xi^{-2} $ (Ref.~\onlinecite{minnhagen:rev}),
leading to $z=2$ for both boundary conditions. 
However, from a computational point of view, the calculation of the size-converged $R$ 
for the FTBC in the high-temperature phase becomes difficult as we approach $T_c$ from above, 
due to the diverging correlation length. On the other hand, if we instead
focus on the scaling of the characteristic frequency $\omega_0$, which
is expected to be proportional to $R$ and can be calculated for the PBC, 
then we do indeed find an indication of 
the expected behavior, $\omega_0 \sim \xi^{-2}$, as seen in 
Fig.~\ref{fig:2drsj_high} for the PBC. 
For the FTBC this result is consistent with our observation
$z \approx 2.0$ at $T_c$, whereas for the PBC the scaling at $T_c$ gives
$z\approx 1.5$, which differs from the expectation. 
Why is there then a difference in the PBC between $z$ values at and above $T_c$?
The point is that the long-time relaxation above $T_c$ is governed by
the thermally created free vortices, 
whose density satisfies $n_F\propto \xi^{-2}$,
whereas the behavior precisely at $T_c$, where $n_F=0$, is instead dominated
by the bound pairs of vortices and antivortices. The conclusion is then that the
constraint imposed by the PBC on the vortex-antivortex escape
gives rise to this peculiar discontinuity of $z$ precisely at $T_c$. This is
in contrast to the FTBC case where $z$ appears to be a continuous function of $T$.

Next we compare the results from the dynamic scaling in equilibrium
and the short-time relaxation method which probes the relaxation 
when the system approaches equilibrium. For RSJD with the FTBC there is no
difference: the resistance scaling and the short-time relaxation method
yield the same $z$ at and below $T_c$ (see Table~\ref{table:2d}). 
However, for RSJD with the PBC the equilibrium scaling and
the short-time relaxation scaling lead
to different results, $z \approx 1.5$ and $z \approx 1.2$, respectively.
In fact by comparing Figs.~\ref{fig:2d_Gt}(a) and \ref{fig:2drsj_short_Tc}(b) 
one realizes that the approach to
equilibrium from the chosen starting nonequilibrium configuration is much slower than
the equilibrium relaxation. Apparently the constraint imposed on
the relaxation by the PBC in combination with the nonequilibrium starting
configuration is causing the difference.

We now turn to the discussions for RD, where for the FTBC we find from the resistance 
scaling the same $z$ at and below $T_c$ as for RSJD (see Table~\ref{table:2d}). 
In this context it is interesting to note that the 2D $XY$ model with the FTBC is 
dual to the lattice CG model with the PBC 
(see Ref.~\onlinecite{olsson} for the mapping between two models),
where the same values of the dynamic critical exponent 
($z = z_{\rm scale} = 1/\tilde{\epsilon}T^{\rm CG}-2$) 
have been found in  MC dynamics.~\cite{weber96} 
Also, the continuum CG model with Langevin dynamics of the pure
relaxational form has been found to give the same values of $z$.~\cite{kenneth}
Accordingly it is tempting to conclude that the result presented in this
work for the 2D $XY$ model with the FTBC is associated with the vortices 
and that it is essential to define the model so as to allow for a
proper relaxation of vortex-antivortex annihilation across the boundary, 
which is not the case for the PBC.
Furthermore, the result that $z = z_{\rm scale}$ appears to be universal
in the sense that it does not matter whether or not the underlying dynamics 
is purely relaxational (such as RD in this work, MC dynamics in Ref.~\onlinecite{weber96}, 
and Langevin dynamics in Ref.~\onlinecite{kenneth}),  or
it has an additional constraint like local current conservation in the RSJD case.

Although the short-time relaxation method applied to RD gives the 
expected value $z\approx 2$ at $T_c$, it fails to yield the equilibrium 
size scaling value below $T_c$.
In addition, if we compare the decay behaviors at and below $T_c$,
shown in Figs.~\ref{fig:2dr_short_Tc} and \ref{fig:2dr_short_low},
respectively, we notice that the time scale of the relaxation does
not depend  significantly on the temperature or on the boundary condition, 
in sharp contrast to RSJD. 
We suggest the following reason: In RD the relaxations of spin waves
and vortices are effectively decoupled and the short-time relaxation
in this case only probes the spin-wave degrees of freedom, 
which follow the purely relaxational dynamics with the trivial exponent $z=2$
at any $T$, while the resistance scaling probes 
the vortex degrees of freedom.
This is then in contrast to RSJD with the FTBC
where both degrees of freedom are strongly coupled,
leading to the same relaxation time (and accordingly the same
$z$ value) for $\tilde{\psi}$ and $R$.
It is also interesting to note that $z\approx 2$ was also found in Ref.~\onlinecite{luo97}
from the MC simulations of the 2D $XY$ model with the PBC at and below $T_c$ by 
using a similar short-time relaxation method.

\subsection{Discussions of the 3D $XY$ model}
We next turn to the 3D $XY$ model (see Table~\ref{table:3d} for a summary of results). 
The discussion for the 2D case in Sec.~\ref{subsec:dis2d} regarding the boundary conditions
carries over to 3D, and we expect that the FTBC has to be used 
whenever the relaxation process associated with the expansion and the subsequent 
annihilation of a vortex loop across the boundary is important because the 
conventional PBC prevents this relaxation. 

For RSJD, $z\approx 1.5$ is found from the linear resistance and the short-time
relaxation method for the FTBC, as well as from the scaling of the supercurrent correlation
at and above $T_c$ for the PBC. In addition, the same value $z \approx 1.5$ 
is also found for RD with the FTBC from the finite-size scaling 
of the linear resistance. 
We note here that the MC simulations of the lattice vortex loop model in 3D 
(Ref.~\onlinecite{weber97}) also have found the same value.
The agreement between the $z$ values for the three different dynamic models
(RSJD and RD with the FTBC, and MC dynamics of the vortex loop model with the PBC)
was also found in 2D.
This value $z\approx 1.5$ obtained in 3D 
is consistent with $z=d/2$ (with $d=3$ in 3D) 
for model E and model F
describing critical dynamics of superfluid systems, in
the classification scheme of Hohenberg and Halperin.~\cite{hohenberg77,noteF}
Consequently, it is again tempting to conclude that the result for the 
3D $XY$ model can be associated with the vortex loops and that the critical dynamics 
of RSJD and RD are equivalent as long as the boundary condition allows for 
the proper vortex loop escape over the boundary. 

As in 2D, we find that the short-time relaxation method for RSJD with the PBC 
(with the result $z \approx 1.2$) does not reflect the true equilibrium relaxation 
corresponding to $z\approx 1.5$, and we  again suggest that  this is due to the
constraint imposed by the PBC. On the other hand, we find that the short-time 
relaxation method for RD with the FTBC gives $z=2.0(1)$ 
[see Figs.~\ref{fig:3d_short_ftbcTc}(b) and \ref{fig:3d_short_ftbc}(b)].
We propose the same explanation as we did for 2D: 
The short-time relaxation in RD at criticality does not reflect the true 
long time relaxation because the vortex loop configurations are still 
out of equilibrium even when $\tilde{\psi}\approx 0$ is reached.
In this respect it is interesting to note that the values $T_c =2.20$
and $\nu = 0.67$ used  in the scaling collapse for $\tilde\psi$ 
in Fig.~\ref{fig:3d_short_ftbc}(b) with $z = 2.0$ agree with the value
expected for the 3D $XY$ model and that the same $T_c = 2.20$ was used in the 
resistance scaling in Fig.~\ref{fig:3d_R}(b) and yielded $z \approx 1.5$.

We next discuss the results for RD with the PBC.
The scalings of the supercurrent correlation both at and above $T_c$ 
(corresponding to the finite-$L$ scaling and the finite-$\xi$ scaling, respectively),
as well as the short-time relaxation at $T_c$,  
consistently give $z\approx 2$. This value corresponds to model A of 
relaxational dynamics in the Hohenberg-Halperin classification 
scheme.~\cite{hohenberg77,dorsey91,wickham99} 
The most striking feature in RD is that the scalings at $T_c$ for the FTBC 
(resistance scaling) and PBC (supercurrent scaling) 
correspond to different values, i.e., $z \approx 1.5$ and $z \approx 2.0$, respectively.

In the high-temperature phase in 3D where $\xi \ll L$, one expects
that $z$ is independent of boundary conditions.~\cite{note3d} 
Consequently, $z \approx 2$ found for RD with the PBC at temperatures
above $T_c$ implies $z\approx 2$ also for RD with the FTBC in the
same high-temperature regime, again consistent with model A.~\cite{hohenberg77} 
In contrast, $z$ determined from the resistance scaling at $T_c$ instead
gives $z\approx 1.5$ for RD with the FTBC. 
We propose the same explanation for this discontinuity of $z$ at $T_c$ 
in the RD case with the FTBC as we did in 2D:
Above $T_c$ where $\xi \ll L$, the finite value of the resistivity 
reflects the physics of dissociated vortex loops 
whereas precisely at $T_c$, where the resistivity vanishes as $L$ is increased,
the physics is dominated by the large nondissociated vortex loops.

We now compare our results in 3D with earlier studies.
Values consistent with $z\approx 1.5$ have also been found in earlier simulations:
$z = 1.5(5)$ was obtained from the $I$-$V$ characteristics
of the current-driven RSJ model with an open boundary (Ref.~\onlinecite{lee92}),
and $z=1.5(1)$ was concluded from the scaling of the linear resistance
for the MC simulations of the $XY$ model in the vortex representation with the PBC
(Refs.~\onlinecite{weber97} and \onlinecite{lidmar98}) 
which corresponds to the FTBC in the phase representation as mentioned above
and explained in Ref.~\onlinecite{olsson}. Finally, MC spin dynamics
applied to the three component $XY$ model gave $z=1.38(5)$ in 
Ref.~\onlinecite{krech99}.
On the other hand, the experimental situation for high-$T_c$ 
superconductors is less clear:~\cite{junod99}
From several zero field dc conductivity experiments 
$z\approx1.5$ has been found on single YBCO-123 crystals~\cite{pomar-holm}
and a similar result $z=1.6(1)$ was also obtained for a Bi-2212 crystal.~\cite{jtkim99b}  
However, from the scaling of the magnetoconductivity of a thick YBCO-123 film 
$z=1.25(5)$ was found in Ref.~\onlinecite{moloni97}, whereas a similar 
experiment reported $z\approx 2$ in Ref.~\onlinecite{jtkim98}.
From a theoretical point of view the renormalization group methods
applied to the relaxational model (model A) yield the result 
$z=2+c\eta$,~\cite{hohenberg77,wickham99} with 
$\eta\approx0.02$ and $c\approx 0.7261$, leading to $z \approx 2.0$. 
However, as far as we know,  no corresponding calculation 
has been made for the 3D RSJ model. One may argue that since
the 3D RSJ model is a {\it bona fide} model of a superconductor the
critical dynamics should belong the dynamic universality 
class of model F which describes superfluids.~\cite{hohenberg77} 
This gives $z=d/2=1.5$ for a model with the static properties given 
by the 3D $XY$ model.~\cite{hohenberg77}

\subsection{Discussions of the 4D $XY$ model}
In case of the 4D $XY$ model both resistance scaling at $T_c$ and short-time 
relaxation give $z\approx 2$ for RSJD as well as for RD
(see Table~\ref{table:4d} for summary of results). 
This is in perfect accordance with the Hohenberg-Halperin
classification scheme where the RSJD case should be related to models E and F
with $z=d/2=2$ and the RD case with the model A value $z=2$. 
This in turn just reflects that 4D is the
upper critical dimension.

\section{Summary} \label{sec:summary}
We have found that the size scaling of the resistance for the $XY$
model with the FTBC gives the dynamic critical exponent $z$ related to
superfluid and superconducting systems with an open boundary.
This is the case in two, three and four dimensions both for relaxational
and RSJ dynamics.
In 2D this applies for $T\leq T_c$, whereas in 3D and 4D the dynamics
is critical only at $T=T_c$.
However, the 3D case with relaxational dynamics has a discontinuity in
the
$z$ value since the relaxation time $\tau$ scales as
$\tau\propto L^{1.5}$ at $T_c$,
whereas it scales as $\tau\propto \xi^{2}$ just above $T_c$. 

The short-time relaxation method, which probes the relaxation from a
nonequilibrium configuration,
does give the same result, except for the 3D case with relaxational
dynamics where
$z\approx 2$ is obtained. This discrepancy shows that although the
short-time
relaxation method very often is reliable and efficient, it cannot
always be trusted
as a determination of the critical equilibrium dynamics.~\cite{bray} 

The $XY$ model with the PBC has a different dynamical size scaling behavior than
with the FTBC. In 2D the $XY$ model with the PBC and RSJ dynamics gives
smaller values of $z$ ($z<2$) both at and below $T_c$. This demonstrates that
the boundary condition influences the size scaling properties of the
dynamics. Also in this case there is a discontinuity in $z$ since  $\tau\propto
L^{1.5}$ at $T_c$
but $\tau \propto \xi^{2}$ just above. This is similar to the
discontinuity found in 3D for
relaxational dynamics.
The short-time relaxation fails to give the equilibrium $z$ at $T_c$ 
for the 2D $XY$ model with the PBC and RSJ dynamics.

In 3D the $XY$ model with RSJ dynamics and the PBC gives the same result
as for the FTBC. Thus in this case the boundary condition does not
influence the size scaling.
This is in contrast to the 3D $XY$ model with relaxational dynamics
which gives different result for the PBC and FTBC. 

In 4D all determinations give the simple relaxational value $z\approx
2$ independent of boundary condition and dynamics.

The actual values determined are consistent with the following
sequences: the $XY$ model with RSJ dynamics and the PBC at $T_c$ is consistent
with $z=1.5$, $1.5$, and $2$ for the 2D, 3D and 4D cases, whereas
the FTBC is consistent with $z=2$, $1.5$, and $2$.
Similarly for relaxational dynamics the PBC gives the sequence $z=2$, $2$,
and $2$ whereas the FTBC gives $z=2$, $1.5$, and $z=2$ for the 2D, 3D,
and 4D cases, respectively.

The $z$ values for a superconductor can be related to the nonlinear
$I$-$V$ exponent $a$ through the scaling relation $a=1+z$.~\cite{fisher91,dorsey91}
The $I$-$V$ measurements correspond to an open boundary and simulations
with the FTBC are consistent with the scaling relation, as shown for the
2D case in Ref.~\onlinecite{bjkim99a}. On the other hand, the $z$ values
calculated with the PBC in 2D do not fulfill the relation
because $a > 1+z$ for $T \leq T_c$.
In this sense, the boundary condition has direct physical significance.

\acknowledgments
The authors thank P. Olsson for useful discussions and
for providing his unpublished high-precision results of 
Monte Carlo simulations.  
This work was supported by the Swedish Natural Research Council
through Contract No. FU 04040-332.

\end{multicols}
\appendix
\section{Scaling form of the supercurrent correlation function} \label{append:scale}

For supercurrent correlation scaling in 2D,
the superfluid density $\rho_s$ is proportional to the vortex
dielectric function $1/\epsilon(0)$ which in turn is related to the
conductivity by $\sigma(\omega)\propto 1/[i\omega\epsilon(\omega)]$.
Precisely at the KT transition $1/\epsilon(0)$ has a logarithmic size
scaling,~\cite{weber-minnhagen}
\[ \frac{1}{\epsilon_{L}(0)}-\frac{1}{\epsilon_{\infty}(0)}\propto
    \frac{1}{\ln(L/c)}\] ,
    where $c$ is a constant. This is consistent with the functional form in
terms of a scaling function with a logarithmic correction for the frequency
dependence 
    \[
    \frac{1}{\epsilon(\omega)}-\frac{1}{\epsilon(0)}\propto\frac{1}{\ln(L/c)}F_\epsilon(\omega \tau) .
        \]
The supercurrent correlation function $G(t)$ is related to $\sigma(\omega)$
by a Fourier transform so that
\[
{\rm Im}\left[\frac{1}{\epsilon(\omega)}\right]=\int_0^\infty dt\omega \cos
\omega t G(t)=\frac{\tilde{F}_\epsilon(\omega \tau)}{\ln L/c},  \]
where $\tilde{F}(x)={\rm Im}[F(x)]$. From this it follows that
\[
\ln(L/c) G(t)\propto \int_{-\infty}^\infty d\omega
\frac{1}{\omega}\tilde{F}_\epsilon(\omega \tau) e^{-i\omega\tau t/\tau}=
\int_{-\infty}^\infty d\omega\tau \frac{1}{\omega\tau}\tilde{F}_\epsilon(\omega
\tau) e^{-i\omega\tau t/\tau}=F_G(t/\tau) ,
\]so that $\ln(L/c) G(t)$ has the scaling form $F_G(t/\tau)$.

In 3D we have instead $1/\epsilon_L(0)\propto\rho_s\propto 1/L$ at $T_c$
and $1/\epsilon_{\infty}(0)=0$ so that this time $1/\epsilon_L(0)-
1/\epsilon_{\infty}(0)\propto 1/L$. This means that going through the
same steps as above for the 3D case gives the scaling form $LG(t)=F_G(t/\tau)$.

\section{Approximation made in the Nyquist formula for linear resistance} \label{append:R}

The step from Eq.~(\ref{eq:R1}) to Eq.~(\ref{eq:Rmu}) is equivalent to showing that
\[
\int_0^{\Theta}dt\langle \dot{\Delta}(t)\dot{\Delta}(0)\rangle
\rightarrow \frac{1}{2\Theta}\langle \left(
  \Delta(\Theta)-\Delta(0)\right)^2 \rangle \]
in the limit of large $\Theta$. The left-hand side can, due to
translational invariance, be written as
\begin{equation}\label{A1}
\frac{1}{\Theta}\int_0^\Theta ds\int_0^{\Theta}dt\langle
  \dot{\Delta}(t+s)\dot{\Delta}(s)\rangle=\frac{1}{\Theta}\int_0^\Theta ds
  \left[\int_{-s}^{\Theta-s}- \int_{-s}^{0}+\int_{\Theta-s}^{\Theta}
\right]dt \langle
  \dot{\Delta}(t+s)\dot{\Delta}(s)\rangle .
\end{equation}
The first term on the right-hand side is 
$\langle (\Delta(\Theta)-\Delta(0))^2\rangle/\Theta$, and 
the second term reduces to
\begin{eqnarray} 
-\frac{1}{\Theta}\int_0^\Theta ds\int_{-s}^0dt\langle
  \dot{\Delta}(t+s)\dot{\Delta}(s)\rangle&=&
  -\frac{1}{\Theta}\int_0^\Theta ds\langle
  [\Delta(s)-\Delta(0)]\dot{\Delta}(s)\rangle\\ 
  &=&
  -\frac{\left\langle \Delta^2(\Theta)-\Delta^2(0) -
  2\Delta(0)\Delta(\Theta)+2\Delta^2(0)\right\rangle}{2\Theta}\\ 
  &=& -\frac{\langle \left[\Delta(\Theta)-\Delta(0)\right]^2\rangle}{2\Theta}, 
 \end{eqnarray}
where we have used $2\Delta(s)\dot{\Delta}(s)=d\Delta^2/ds$. 
Thus the sum of the first two terms on the right-hand side 
of Eq.~(\ref{A1}) is equal to 
$\langle \left[\Delta(\Theta)-\Delta(0)\right]^2\rangle/2\Theta$ 
and it remains to prove that the third term vanishes 
in the limit $\Theta\rightarrow\infty$.
This can be realized by changing the order of integration:
\begin{eqnarray}
\frac{1}{\Theta}\int_0^{\Theta}ds\int_{-s}^0dt\langle
\dot{\Delta}(\Theta+t+s)\dot{\Delta}(s)\rangle &=&
\frac{1}{\Theta}\int_{-\Theta}^0dt\int_{-t}^{\Theta}ds\langle
\dot{\Delta}(\Theta+t)\dot{\Delta}(0)\rangle\\ &=&
\frac{1}{\Theta}\int_{-\Theta}^0dt(\Theta+t)\langle
\dot{\Delta}(\Theta+t)\dot{\Delta}(0)\rangle\\&=&
\frac{1}{\Theta}\int_{0}^{\Theta}dxx\langle
\dot{\Delta}(x)\dot{\Delta}(0)\rangle .
\end{eqnarray}
A finite relaxation time $\tau$ means that $\langle
\dot{\Delta}(t)\dot{\Delta}(0)\rangle\propto \exp(-t/\tau)$,
which means that the last integral is finite for any finite
$\tau$. This is the case at $T_c$ whenever the system size is
finite since $\tau\propto L^z$. Consequently, the third term
vanished in the limit $\Theta\rightarrow\infty$ for any finite
$L$ and the step from Eq.~(\ref{eq:R1}) to Eq.~(\ref{eq:Rmu}) follows.

\begin{multicols}{2}

\narrowtext
\vskip 1cm

\begin{table}
\caption{Dynamic critical exponent $z$ for 2D RSJD and RD with the PBC and FTBC. }
\begin{tabular}{l c c c c}
&\multicolumn{2}{c}{RSJD }& \multicolumn{2}{c}{RD}\\
     &    PBC   &        FTBC  &    PBC       &   FTBC   \\ \hline
at $T=0.90$            &     &          &        &     \\ 
resistance scaling     &    $-$         &   2.0(1)   &   $-$         &   2.0(1)    \\
supercurrent scaling   &    1.5(1)         &   $-$   &   2.0(1)         &   $-$    \\
short-time relaxation  &    1.2(1)         &   2.0(1)   &   2.0(1)         &   2.0(1)    \\  \hline
at $T=0.80$            &     &          &        &    \\
resistance scaling     &    $-$         &   3.3(1)   &   $-$         &   3.3(1)    \\
short-time relaxation  &    1.4(1)         &   3.2(1)   &    2.0(1)         &   2.0(1)    \\ \hline
above $T_c$            &     &          &        &    \\
from $\omega_0 \sim \xi^{-z}$  &     $\sim 2$      &  $-$ &  $2.0^{\rm a}$ & $-$     \\
\end{tabular}
$^{\rm a}$ Reference~\onlinecite{jonsson}
\label{table:2d}
\end{table}

\begin{table}
\caption{Dynamic critical exponent $z$ for 3D RSJD and
RD with the PBC and FTBC.}
\begin{tabular}{l c c c c}
&\multicolumn{2}{c}{RSJD }&
\multicolumn{2}{c}{RD}\\
            &    PBC   &       FTBC  &    PBC   &     FTBC                  \\ \hline
resistance scaling     &    $-$   &        1.46(6)   &   $-$        &  1.5(1) \\
supercurrent scaling ($L$)  &    1.5(1) &      $-$   &    2.0(1)&  $-$      \\
supercurrent scaling ($\xi$)  & 1.4(1)    &   $-$   &  1.9(2)    &    $-$      \\
short-time relaxation  &    1.2(1) &    1.50(5)   &   2.1(1)    &    2.0(1) \\
\end{tabular}
\label{table:3d}
\end{table}
\begin{table}
\caption{Dynamic critical exponent $z$ for 4D RSJD and
RD both with the FTBC.}
\begin{tabular}{l c c}
 & RSJD & RD \\ \hline
resistance scaling     &    $\sim 2$  &  $\sim 2$  \\
short-time relaxation  &    2.0(1) &    2.0(1) \\
\end{tabular}
\label{table:4d}
\end{table}

\newpage

\end{multicols}

\begin{thebibliography}{10}
\bibitem{check}
For RSJD at $T=1.40$ with $L=64$, $R \approx 0.44$ and $R \approx 0.425$
were obtained for the FTBC and PBC, respectively.
\bibitem{noteF}
Model F gives $z=3/2$ in 3D when the specific heat exponent is less or equal
  to zero, which is the case for the 3D $XY$ model.
\bibitem{note3d} Explicitly checked and confirmed in 2D (see Ref.~\onlinecite{check}), but 
should apply also in 3D.
\end{thebibliography}
\end{document}